%% file: main.tex
\documentclass[sigconf, nonacm]{acmart}

\usepackage{xspace}
\usepackage{booktabs} 
\usepackage[english]{babel}
\usepackage{xcolor}
\usepackage{xspace}
\usepackage{colortbl}
\usepackage{pifont}
\usepackage{subfigure}
\usepackage{graphicx}
\usepackage{booktabs}
\usepackage{threeparttable}

\usepackage{listings}
\usepackage{float}
\usepackage{multirow}
\usepackage{fancybox} 
\usepackage{amsmath}
\usepackage{hhline}
\usepackage{amsthm}
\usepackage{comment}
\usepackage{tabularx}
\usepackage{bbm}
\usepackage{amsfonts}
\usepackage{pythonhighlight}
\usepackage[inline,shortlabels]{enumitem}
\usepackage{amsthm} 
\usepackage{tcolorbox}
\tcbuselibrary{theorems}

\newcommand\vldbdoi{XX.XX/XXX.XX}
\newcommand\vldbpages{XXX-XXX}
\newcommand\vldbvolume{19}
\newcommand\vldbissue{9}
\newcommand\vldbyear{2026}
\newcommand\vldbauthors{\authors}
\newcommand\vldbtitle{\shorttitle} 
\newcommand\vldbavailabilityurl{}
\newcommand\vldbpagestyle{empty}

\lstdefinelanguage{Rust}{
  morekeywords={
    as, break, const, continue, crate, else, enum, extern, false, fn,
    for, if, impl, in, let, loop, match, mod, move, mut, pub, ref,
    return, self, Self, static, struct, super, trait, true, type, unsafe,
    use, where, while, dyn, async, await, abstract, become, box, do,
    final, macro, override, priv, typeof, unsized, virtual, yield, try
  },
  sensitive=true,
  morecomment=[l]{//},
  morecomment=[s]{/*}{*/},
  morestring=[b]{"},
  alsoletter={:},
}

\usepackage{xcolor}
\definecolor{redorange}{rgb}{1.0, 0.07, 0.0}  

\definecolor{compasscyan}{RGB}{70,181,170}
\definecolor{compassgreen}{RGB}{35, 139, 69}
\definecolor{videostormblue}{RGB}{44,104,245}
\definecolor{hunter_deepgreen}{RGB}{0,150,0}

\definecolor{boxcolor}{gray}{0.9}

\usepackage[linesnumbered,ruled,noend,scleft]{algorithm2e}
\usepackage{float}
\usepackage{algorithmicx}

\definecolor{cardinal}{rgb}{0.77, 0.12, 0.23}

\SetCommentSty{mycommfont}

\newenvironment{denseitemize}{
\begin{itemize}
}{\end{itemize}}

\newenvironment{denseenum}{
\begin{enumerate}
}{\end{enumerate}}

\definecolor{codegreen}{rgb}{0,0.6,0}
\definecolor{codegray}{rgb}{0.5,0.5,0.5}
\definecolor{codepurple}{rgb}{0.58,0,0.82}
\definecolor{backcolour}{rgb}{0.95,0.95,0.92}
\definecolor{darkpurple}{RGB}{157,0,255}

\definecolor{eclipseStrings}{RGB}{42,0.0,255}
\definecolor{eclipseKeywords}{RGB}{127,0,85}
\colorlet{numb}{magenta!60!black}

\lstdefinestyle{codestyle}{
    keywordstyle=\color{black}\bfseries,
    numberstyle=\color{codegray},
    basicstyle=\ttfamily\mdseries\scriptsize,
    emph={let},
    emphstyle={\color{black}\bfseries},
    breakatwhitespace=false, 
    frame=lines,      
    rulecolor=\color{codegray},  
    breaklines=true,                 
    captionpos=b,                    
    keepspaces=true,                 
    numbers=left,                    
    numbersep=5pt,                  
    showspaces=false,                
    showstringspaces=false,
    showtabs=false,                  
    tabsize=2
}
\definecolor{eclipseStrings}{RGB}{42,0.0,255}
\definecolor{eclipseKeywords}{RGB}{127,0,85}
\colorlet{numb}{magenta!60!black}
\lstdefinelanguage{json}{
    basicstyle=\scriptsize\ttfamily,
    commentstyle=\color{eclipseStrings}, 
    stringstyle=\color{eclipseKeywords}, 
    emph={AdaEmbed,emb_agent},
    emphstyle={\color{black}\bfseries},
    numbers=left,
    numberstyle=\scriptsize,
    stepnumber=1,
    numbersep=8pt,
    showstringspaces=false,
    breaklines=true,
    frame=lines,
    string=[s]{"}{"},
    comment=[l]{:\ "},
    morecomment=[l]{:"},
    literate=
        *{0}{{{\color{numb}0}}}{1}
         {1}{{{\color{numb}1}}}{1}
         {2}{{{\color{numb}2}}}{1}
         {3}{{{\color{numb}3}}}{1}
         {4}{{{\color{numb}4}}}{1}
         {5}{{{\color{numb}5}}}{1}
         {6}{{{\color{numb}6}}}{1}
         {7}{{{\color{numb}7}}}{1}
         {8}{{{\color{numb}8}}}{1}
         {9}{{{\color{numb}9}}}{1}
}

\def\ie{{i.e.}}
\def\eg{{e.g.}}

\def\inline1x{Model-S}

\def\name{Compass\xspace}
\usepackage{tikz}
\newcommand*\circled[1]{\tikz[baseline=(char.base)]{
            \node[shape=circle,draw,inner sep=0pt] (char) {#1};}}

\algnewcommand{\LeftComment}[1]{\Statex \(\triangleright\) #1}

\newcommand{\todo}[1]{}
\newcommand{\remark}[1]{}
\newcommand{\banruo}[1]{}
\newcommand{\fan}[1]{}
\newcommand{\hunter}[1]{}
\newcommand{\revision}[1]{#1}

\newcommand{\bipar}[1]{  \paragraph{\bfseries\itshape #1}}

\newtheorem{theorem}{Theorem}

\newtcolorbox{highlighttheorem}{
    colback=white!95!green,
    colframe=teal!60!black,
    arc=2pt,
    boxsep=5pt,
    left=1pt,
    right=10pt,
    top=2pt,
    bottom=2pt,
    boxrule=0.8pt
}

\begin{document}

\hypersetup{
  linkcolor=videostormblue, 
  citecolor=compassgreen,    
  filecolor=black,       
  urlcolor=black          
}

\title{\name: SLO-aware Query Planner for Compound AI Serving at Scale}

\input{sections/abstract}

\author{Banruo Liu}
\affiliation{%
  \institution{University of Illinois Urbana-Champaign}
  \city{Urbana, IL}
}
\email{banruol2@illinois.edu}

\author{Wei-Yu Lin}
\affiliation{%
  \institution{Unaffiliated}
  \city{Urbana, IL}
}
\email{weiyulin68@gmail.com}

\author{Minghao Fang}
\affiliation{%
  \institution{University of Illinois Urbana-Champaign}
  \city{Urbana, IL}
}
\email{mf46@illinois.edu}

\author{Yihan Jiang}
\affiliation{%
  \institution{University of Illinois Urbana-Champaign}
  \city{Urbana, IL}
}
\email{hunterj3@illinois.edu}

\author{Fan Lai}
\affiliation{%
  \institution{University of Illinois Urbana-Champaign}
  \city{Urbana, IL}
}
\email{fanlai@illinois.edu}

\maketitle

\pagestyle{\vldbpagestyle}
\begingroup\small\noindent\raggedright\textbf{PVLDB Reference Format:}\\
\vldbauthors. \vldbtitle. PVLDB, \vldbvolume(\vldbissue): \vldbpages, \vldbyear.\\
\href{https://doi.org/\vldbdoi}{doi:\vldbdoi}
\endgroup
\begingroup
\renewcommand\thefootnote{}\footnote{\noindent
This work is licensed under the Creative Commons BY-NC-ND 4.0 International License. Visit \url{https://creativecommons.org/licenses/by-nc-nd/4.0/} to view a copy of this license. For any use beyond those covered by this license, obtain permission by emailing \href{mailto:info@vldb.org}{info@vldb.org}. Copyright is held by the owner/author(s). Publication rights licensed to the VLDB Endowment. \\
\raggedright Proceedings of the VLDB Endowment, Vol. \vldbvolume, No. \vldbissue\ %
ISSN 2150-8097. \\
\href{https://doi.org/\vldbdoi}{doi:\vldbdoi} \\
}\addtocounter{footnote}{-1}\endgroup

\ifdefempty{\vldbavailabilityurl}{}{
\vspace{.3cm}
\begingroup\small\noindent\raggedright\textbf{PVLDB Artifact Availability:}\\
The source code, data, and/or other artifacts have been made available at \url{\vldbavailabilityurl}.
\endgroup
}

\input{sections/introduction}
\input{sections/background}

\input{sections/overview}
\input{sections/design}
\input{sections/implementation}

\input{sections/evaluation}
\input{sections/related}
\input{sections/conclusion}

\label{EndOfPaper}

\bibliographystyle{ACM-Reference-Format}
\bibliography{main}
\clearpage

\appendix
\input{appendix/appendix}

\end{document}

%% file: sections/abstract.tex
\begin{abstract}
The rise of compound AI serving that integrates multiple operators in a pipeline enables end-user applications such as generative AI-powered meeting companions, autonomous driving, and immersive gaming. These workloads span diverse deployment spaces, from cloud-only queries to edge-assisted ones across infrastructure tiers, often including both within an application. Achieving high service goodput---i.e., meeting service level objectives (SLOs) for pipeline latency, accuracy, and costs---requires joint planning of operators' placement, configuration, and resource allocation. However, diverse SLOs, varying runtime environments (e.g., heterogeneous device speeds), and a large volume of queries competing for shared infrastructure explode the planning space, making real-time serving and cost-efficient deployment intractable with existing advances.

This paper presents Compass, the first SLO-aware query planner that optimizes large-scale compound AI workloads across diverse deployment spaces. Compass decomposes the many-query, multi-SLO planning problem into tractable subproblems while preserving global decision quality, exploiting plan similarities within and across queries to slash the search steps.
It further improves per-step efficiency with a plan profiler that performs selective profiling to achieve high-fidelity performance estimates at a fraction of the profiling cost. At runtime, Compass performs query-plan bipartite matching to maximize SLO goodput under resource contentions. Real-world evaluations show that Compass improves service goodput by 2.4--5.1$\times$,  reduces deployment costs by 3.8--4.5$\times$, and accelerates planning by 4.2--10.5$\times$, achieving service responsiveness within seconds and near-optimal decision quality.
\end{abstract}

%% file: sections/introduction.tex
\section{Introduction}
\label{sec:intro}

\begin{figure*}[t]
    \centering
    \includegraphics[width=0.8\linewidth]{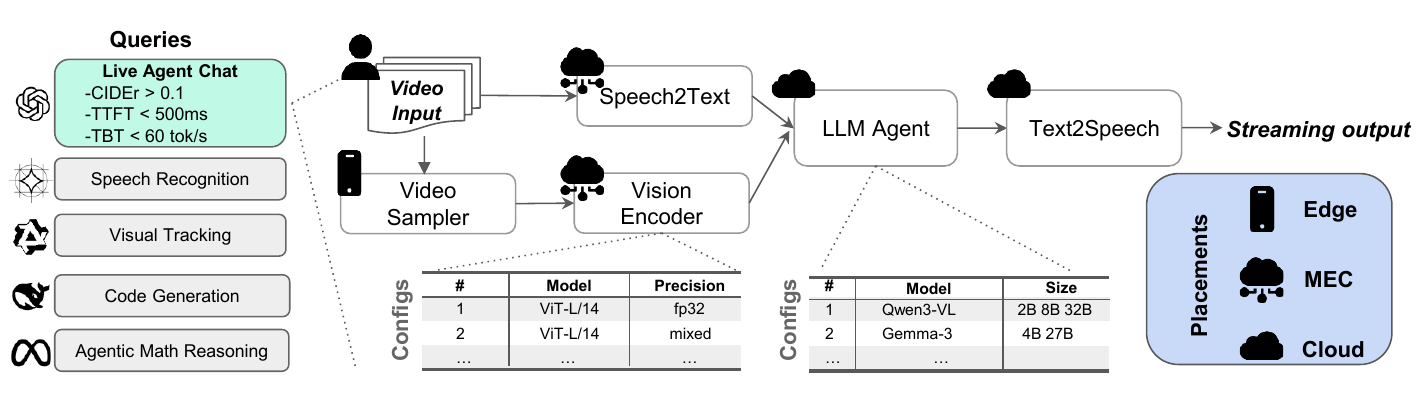}
    \caption{As compound AI services increasingly cater to end users, their deployment can span multi-tier, heterogeneous infrastructure.}
    \label{fig:compound_ai_workflow}
\end{figure*}

Machine learning (ML) serving is shifting from deploying monolithic models to composing pipelines of multiple operators, such as video denoisers, person and pose detectors, feature extractors, large language model (LLM)-based reasoning modules, and speech and avatar renderers for a live video chat query with AI agents like \textit{``track and imitate user's body movements while performing push-ups, provide instant corrective feedback on muscle engagement''}. By adding and distributing augmented operators, compound AI serving enables better accuracy~\cite{recall-arxiv24}, access control (e.g., via local data processing~\cite{oort-osdi21}), efficiency (e.g., reduced data transfer~\cite{soc-atc24}), and operational flexibility (e.g., autoscaling). 
Increasingly, compound AI applications span diverse use cases, including AI companions in live video conferences~\cite{zoom-aicompanion}, ChatGPT's live video conversation~\cite{chatgpt-videochat, chatgpt-livevideo}, speech-to-text streaming~\cite{openai-tts}, autonomous driving~\cite{emp-mobicom21},  and immersive gaming~\cite{gcp-genai-gaming}, with deployments across cloud machines or even infrastructure tiers with varying capabilities (e.g., from edge devices and near-edge clusters to the cloud~\cite{zhang2024vulcan}). 

As with many advances in serving single ML models~\cite{zhang2023shepherd, clockwork-osdi20, andes-arxiv24, sarathi-osdi24}, maximizing the number of queries meeting their service level objectives (\ie, SLO goodput) is critical for both service providers and users.  
However, compound AI serving poses unique challenges: query planning must holistically consider the \underline{\emph{configuration}} of operators (\eg, video sampling rates, Qwen3 vs. Llama3 agents), their physical \underline{\emph{placements}} (\eg, cloud-machine selection and edge offload choices that reduce intermediate traffic), and \underline{\emph{resource allocation}} (\eg, GPU or bandwidth provisioning). 
These decisions directly affect SLO metrics such as latency, accuracy, and cost, making query planning a fundamental systems challenge~\cite{compoundai-blog}.

However, SLO requirements vary across applications~\cite{chatgpt-videochat, basalla2021latencyofecommerce, agarwal2024tarzan-videoconferencing} and even among users of the same application (e.g., free vs. paid users, or varying speaking speeds in GenAI chat~\cite{andes-arxiv24}).
Worse yet, pipelines with identical operators can exhibit performance variations due to heterogeneous input (\eg, accuracy changes under different lighting conditions~\cite{assistive-ai}) and system capabilities (\eg, edge-cloud network bandwidth). 
Moreover, real deployments face high query volumes (\eg, tens of thousands of concurrent ChatGPT live video chat sessions~\cite{chatgpt-livevideo}), resource-intensive operators, and dynamic runtime environments. These factors make it essential to plan many pipelines efficiently, minimizing deployment costs under resource contention while sustaining SLOs in real time (\S\ref{sec:background}).


Unfortunately, the design space of existing ML query planners remains fragmented, targeting either single, long-running queries~\cite{zhang2024vulcan, wu2022jellybean} or homogeneous cloud-only environments~\cite{zhang2017videostorm} (Table~\ref{tab:design-space}). 
Yet, compound AI applications span diverse deployment spaces, with queries spreading both cloud-only and edge-cloud tiers to maximize SLO goodput (\S\ref{sec:challenges}).
Worse, they falter under the combinatorial explosion of candidate plans in compound AI pipelines. For example, an agentic video chat query can yield thousands of candidate plans, involving various placements, configurations, and resource allocations. 
\revision{Planning overheads of existing advances can stretch to tens of minutes, taking 40\%-62\% total time including serving, becoming the major bottleneck,
outlasting the lifetime of many queries and making real-time serving infeasible,
while still incurring substantial SLO violations and inflated costs (\S\ref{sec:opportunity}).}

We introduce \name (\S\ref{sec:overview}), the first SLO-aware query planner for \underline{Comp}ound \underline{A}I \underline{s}erving at \underline{s}cale, capable of supporting diverse deployment spaces (e.g., cloud-only, multi-tier, and multi-query settings). 
It simultaneously serves the needs of three key stakeholders: (i) high SLO attainment and real-time responsiveness for users, (ii) cost-efficient deployment for service providers,
and (iii) high resource utilization for infrastructure operators. 
It novelly decomposes the many-query, multi-SLO planning problem into efficient single-query planning, and performs query-plan bipartite matching (combination) under contention to maintain global decision quality.

\revision{
Searching for a cost-efficient plan, even for a single query, is challenging due to the vast search space, multi-dimensional SLOs (e.g., accuracy, latency, and cost), and the heterogeneous plan profiling costs in each search step. For example, profiling a plan's final accuracy that involves LLaMA3-70B inference is 9$\times$ more expensive, in both GPU time and cost, than that with LLaMA3-8B. 
To minimize search steps, we design a search optimizer that decomposes multi-SLO objectives into separate surrogate optimizers, maximizing knowledge reuse across similar plans (e.g., defined by their performance proximity in the learned latent feature space), both within and across queries (\S\ref{sec:search-optimizer}). The optimizer prioritizes high-potential candidates at each step while accounting for heterogeneous profiling costs.
To cut per-step overhead, we introduce a profiler that adaptively adjusts profiling amount, reuses prefix caches from prior runs, saving cost while ensuring provable profiling correctness (\S\ref{sec:profiler}).}

Maximizing SLO service goodput, however, requires joint planning across many concurrent queries, intensifying the combinatorial explosion.
Planning queries independently ignores resource contention in shared infrastructure, yielding globally inefficient allocations and SLO violations (e.g., each query's optimal plan may prefer the cloud tier, leaving edge tiers idle). 
\name addresses this with an integral two-stage approach that preserves global decision quality: 
the single-query planner first identifies the Pareto-optimal set of candidate plans that meet each query's SLOs. Then, 
\name efficiently performs bipartite matching between the query set and their Pareto-optimal set of plan candidates, selecting query-plan combinations that maximize SLO goodput (\S\ref{sec:multi-query}). 

\name is API-compatible with Kubernetes~\cite{k8s}, supporting existing ML serving stacks with a few lines of code change (\S\ref{sec:implementation}).
Evaluations across realistic cloud-only and edge-cloud deployments---spanning five representative compound AI applications such as live agentic video chat and code generation---show that \name delivers 2.4--5.1$\times$ higher service goodput than existing advances~\cite{zhang2024vulcan, zhang2017videostorm}, even in their targeted deployment space (\eg, cloud-only setting). Furthermore, \name reduces deployment costs by 3.8--4.5$\times$, and accelerates planning by 4.2--10.5$\times$, \revision{enabling second-level query responsiveness and \underline{\emph{near-optimal}} decision quality across diverse deployment spaces (\S\ref{sec:eval}).}

\revision{
Overall, we make the following contributions in this paper:
\begin{denseitemize}
    \item We develop the first generic SLO-aware query planner supporting compound AI serving at scale across deployment spaces.
    
    \item We propose a novel search mechanism that effectively decomposes global planning, leveraging plan similarities and adaptive profiling to navigate multi-SLO planning.
    
    \item We evaluate \name in various real-world settings, showing significant SLO improvements and cost savings.

\end{denseitemize}
}

%% file: sections/background.tex
\section{Background and Motivation}
\label{sec:background}

We begin with an overview of compound AI (\S\ref{sec:compound-ai}), followed by the challenges it faces (\S\ref{sec:challenges}), and conclude with the limitations of current advances that motivate our work (\S\ref{sec:opportunity}).

\subsection{Compound AI Serving}
\label{sec:compound-ai}

Compound AI serving relies on multi-stage pipelines to generate the final output. For example (Figure~\ref{fig:compound_ai_workflow}), a video chat with the assistive AI agent may require operators that sample video frames, encode images, and then invoke a multi-modality model~\cite{videounderstanding-iclr25}, and convert text output back to audio. Each operator typically exposes multiple configuration options (e.g., frame sampling rates and model sizes). 

These configurations interact with placement and resource allocation decisions. Cloud machines are capable yet expensive and easily overloaded due to the prevalence of load burstiness, while edge devices are resource-constrained yet free and close to the data source. Near-edge clusters (e.g., MEC~\cite{azuremec}) sit in between, providing moderate cost and moderate capacity as an attractive offload layer. We show that using multi-tier resources can significantly improve service goodput (\S~\ref{eval:perf-breakdown}). However, placement decision of each operator becomes complicated; for example, placing operators closer to the data source (e.g., on end devices) can reduce traffic over the Internet but may increase compute latency, ultimately impacting the pipeline's overall latency and accuracy (\S\ref{sec:challenges}).

\begin{table*}[t]
  \centering
  \small

  \begin{tabular}{l|c|c|c|c|l|l}
      \hline
        \textbf{System} & \textbf{Multi-query} & \textbf{Multi-tier} & \textbf{Heter. Aware.} & \textbf{Contention Aware.} & \textbf{Response Time} & \textbf{Near-Optimal Time }\\
      \hline      
      Exhaustive Search & \checkmark & \checkmark & \checkmark & \checkmark  & > 2 hours & > 24 hours \\
      Videostorm~\cite{zhang2017videostorm} & \checkmark & $\times$ & $\times$ & \checkmark  & $\sim$1.6 min & $\sim$34 min \\
      Vulcan~\cite{zhang2024vulcan} & $\times$ & \checkmark  & \checkmark & $\times$ & $\sim$1.2 min & $\sim$7 min \\
      \textit{Compass} & \checkmark & \checkmark & \checkmark & \checkmark & \textit{< 5s} & \textit{< 20s} \\ 
      \hline
  \end{tabular}
  \caption{\name enables real-time planning for compound AI workloads across deployment spaces. Unlike prior systems that take minutes to hours, \name finds the first SLO-compliant plan (``Response time.'') in under 5s and near-optimal plans (``Near-Optimial.'') in under 20s (\S\ref{eval:e2e}), while comprehensively supporting multiple features. Such speed is essential for powering user-facing services~\cite{gemini-videochat, openai-tts, paraserve-arxiv25, applunch-fast23}.}
  \label{tab:design-space}
\end{table*}

\bipar{Design Space.}
Satisfying SLOs demands an efficient query planner to identify the optimal configuration, placement, and resource allocation for each operator. Practical compound AI queries primarily fall into two categories: 
\begin{denseitemize} 
\item \emph{Queries with Continuous Interaction}: \revision{
These involve continuous input streams, such as assistive video AI agents (often lasts minutes)~\cite{assistive-ai, chatgpt-videochat}, AI-powered mobile cloud gaming~\cite{gcp-aigame}, and video conferencing~\cite{zoom-aicompanion}. They require fast planning to launch and adapt real-time services under dynamic conditions~\cite{applunch-fast23}, while identifying cost-effective plans to scale for many queries and users. Even throughput-intensive workloads, like offline video captioning~\cite{focus-osdi2018}, can strain current planners, which often require hours of search (\S\ref{eval:e2e} and Appendix~\ref{app:eval_results}) and demand minimizing costly GPU-based accuracy profiling for each configuration. }

\item \emph{Queries with Ephemeral Interaction}: 
These pipelines handle numerous short-lived requests, such as  speech recognition or agentic code generation from many users, and aim to optimize average serving performance. Diverse applications, SLO tiers (e.g., free vs. paid users~\cite{apidog2023chatgpt-free-vs-paid}), and runtime dynamics---\eg, changes in input distributions, resource availability, and bursty arrivals---require real-time, cost-efficient (re-)planning to sustain performance.
\end{denseitemize}

As shown in Table~\ref{tab:design-space}, with the growing diversity and scale of applications and users, queries from different or within applications increasingly span deployment spaces for efficiency, cost-effectiveness, and data governance. 
For example, video understanding queries may process live chat streams directly from end-user devices (e.g., live assistive AI video agents~\cite{assistive-ai}), distributing operators across edge, near-edge, and cloud infrastructure \cite{chatgpt-videochat, zoom-aicompanion}, or may run entirely on offline video corpora in the cloud~\cite{focus-osdi2018}. This widening operational footprint makes a unified, cross-space query planner indispensable.

\subsection{Challenges in Compound AI Serving}
\label{sec:challenges}

While optimizing SLO service goodput (i.e., cost-effectiveness) remains the key as prior advances for single ML models~\cite{zhang2023shepherd, infaas-atc21}, compound AI serving introduces unique challenges due to its wide span of operators and deployment scenarios.

\begin{figure}[t]
  \centering
    {
    \subfigure[Heterogeneous Data. \label{fig:heter-data}]{\includegraphics[width=0.48\linewidth]{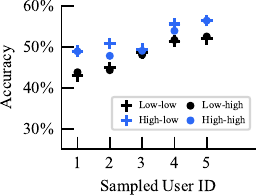}} 
    \hfill
    \subfigure[Vast search space. \label{fig:vast-space}]{\includegraphics[width=0.48\linewidth]{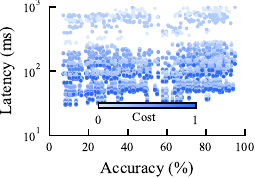}}
  }
  \caption{(a) Input distributions vary across users, causing accuracy variance for the same plan. ``Low-high'': resource-light configurations for operator 1 and resource-intensive configurations for operator 2. (b) Each query can have thousands of plans.}
  \label{fig:model-tradeoff}
\end{figure}

\bipar{Diverse SLO Requirements and Runtime Environments.}
SLO requirements vary not only across applications but also among users within the same application~\cite{infaas-atc21, apidog2023chatgpt-free-vs-paid, agarwal2024tarzan-videoconferencing}. For example, speaking speeds differ by language and age in live agent chat~\cite{andes-arxiv24}; user playstyles and reaction times vary widely in AI-powered gaming~\cite{gaming-speed}; and users interacting with ChatGPT's live video chat for real-world assistance exhibit varying responsive needs~\cite{assistive-ai}.

Even under identical SLO targets, queries often encounter heterogeneous runtime environments and input data characteristics. Operators may be distributed across machines with varying capabilities for cost efficiency (e.g., mixing high- and low-end GPUs in the cloud) or even span heterogeneous tiers such as edge devices, MEC, and cloud platforms (e.g., for live edge inputs). At the same time, input distributions are query- and user-specific. For example, our experiments on real-world video chats~\cite{videommeds} (Figure~\ref{fig:heter-data}) reveal that the accuracy of a fixed pipeline varies widely depending on the user's video input, such as content and lighting conditions~\cite{assistive-ai}. Importantly, more resource-intensive configurations do not always yield higher accuracy, making pipeline planning intractable. All these necessitate customized pipelines. 

\bipar{Enormous Plan Search Space and Multi-query at Scale.}
The interplay of diverse SLOs, heterogeneous system speeds, and data characteristics results in a vast search space. For a pipeline with $M$ operators, each with $N$ configuration options and $K$ placement choices---where $K = 1$ under homogeneous machines, but cost-optimal deployments often prefer heterogeneous machines, thus placements, even within a tier~\cite{compoundai-blog}---the number of plan candidates grows exponentially, O($K^M \cdot N^M$), yielding thousands of candidates (Figure~\ref{fig:vast-space}).  

Moreover, practical deployments must serve large volumes of queries that share underlying infrastructure resources---e.g., today's live assistive AI video application supporting thousands of users concurrently~\cite{chatgpt-livevideo,  zoom-aicompanion}---to maximize utilization and SLO goodput. This requires contention-aware joint planning (e.g., co-locating pipelines on the same machine), which further expands the planning space far beyond single-query scenarios. Worse, bursty workloads and performance fluctuations (\eg, network fluctuations) complicate trade-offs between SLO compliance and deployment cost.

\subsection{Limitations of Existing Advances}
\label{sec:opportunity}

Existing ML serving systems have focused on optimizing either server-side metrics, such as serving throughput~\cite{sarathi-osdi24, alpaserve-osdi23, focus-osdi2018}, or SLO-aware request scheduling for individual ML models~\cite{zhang2023shepherd, distserve-osdi24}. As summarized in Table~\ref{tab:design-space}, a few recent works target live ML analytics (\eg, video analytics), but are fundamentally limited to single, long-running pipelines~\cite{zhang2024vulcan, wu2022jellybean} and/or homogeneous cloud deployments~\cite{zhang2017videostorm}. Beyond their piecemealed support, we next show how they fail to meet the demands of practical compound AI serving.

\begin{figure}[t]
  \centering
    {
    \subfigure[Time to first feasible plan. \label{fig:ttffp-cdf}]{\includegraphics[width=0.48\linewidth]{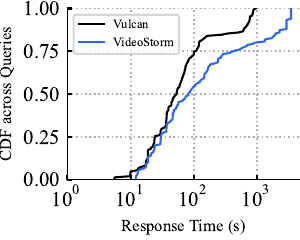}}
    \hfill 
    \subfigure[Global planning is needed. \label{fig:local-global}]{\includegraphics[width=0.48\linewidth]{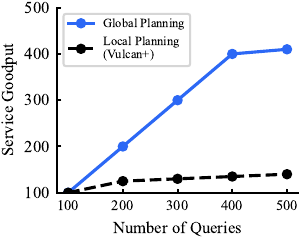}}
  }
  \caption{(a) Existing advances, Vulcan~\cite{zhang2024vulcan} and VideoStorm~\cite{zhang2017videostorm}, spend tens of seconds to identify the first feasible plan that satisfies SLO; and (b) are fundamentally limited for multi-query planning.}
\end{figure}

\bipar{Infeasibility for Single-Query Deployment.} 
Existing efforts suffer from slow responsiveness and high costs in planning, even in their targeted single-query planning scenario. Figure~\ref{fig:vast-space} shows that an agentic video chat query can have thousands of possible plans with widely varying performance and deployment costs. Navigating this explosive search space, existing advances such as Vulcan~\cite{zhang2024vulcan} and VideoStorm~\cite{zhang2017videostorm} take minutes to identify the first SLO-compliant plan (Figure~\ref{fig:ttffp-cdf}), long before any effort to optimize for the plan's deployment cost. 
\revision{
Such sluggish responsiveness renders real-time or short-lived queries infeasible---e.g., brief speech recognition queries often lasts only a dozen seconds (Table~\ref{tab:total-serving})---and prevents rapid replanning to sustain services under runtime dynamics~\cite{openai-tts}. 
Even for queries without strict responsiveness demands (\eg, offline video captioning), they cause exploring large plan spaces, imposing extensive, costly GPU-based plan profiling (\S\ref{eval:e2e}).}

\bipar{Inability for Multi-Query Deployment.} 
In practical settings with many queries, existing approaches incur substantial SLO goodput loss due to their limited design scope. For example, we extend Vulcan to support multi-query deployment by allocating one hour per query to identify its local optimal plan and pack queries across machines. Even with this augmentation and highly relaxed responsiveness requirements, the SLO goodput of Vulcan+ remains far below that of an oracle performing joint global planning (Figure~\ref{fig:local-global}).
\revision{This gap arises because single-query planners ignore cross-query resource contention: when multiple queries share infrastructure, (1) latency deviations caused by contention invalidate previously selected local plans, triggering widespread SLO violations; and (2) per-query optimal plans may overload one tier while leaving others idle (e.g., edge devices). Addressing this, without global multi-query planning, can require aggressive resource overprovisioning, which further reduces overall efficiency, especially under runtime dynamics.
Unfortunately, global planning is intractable, with the exhaustive searching time ballooning to many hours for a dozen queries (\S\ref{eval:e2e}). Such planning latency makes it impractical to scale online deployments, co-locate queries, or adapt to runtime changes in a timely manner.}

%% file: sections/overview.tex
\section{\name Overview}
\label{sec:overview}

We next introduce \name, the first SLO-aware query planner for compound AI serving that enables real-time planning across diverse deployment spaces, including cloud-only, multi-tier, and settings where these coexist with multiple queries. \name simultaneously achieves high SLO attainment and fast responsiveness for users, cost-efficient deployments for service providers, and maximized resource utilization for infrastructure operators.

\begin{figure}
    \centering
    \includegraphics[width=0.9\linewidth]{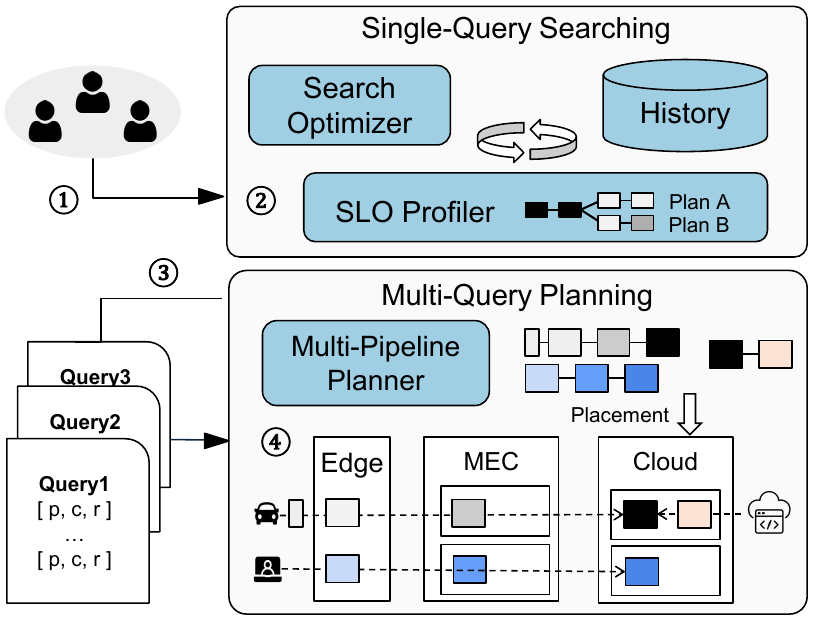}
    \caption{\name overview for compound AI serving at scale.}
    \label{fig:overview}
\end{figure}

\bipar{System Components.}
In practical deployments, queries arrive online and exit once completed. 
For each query, \name determines the query plan, specifying operator placement, configuration, and resource allocation, where it accounts for resource contention and may collocate pipelines (e.g., sharing machines or models). The resource orchestrator (e.g., Kubernetes~\cite{k8s}) then deploys the plan across machines and potentially tiers. 
\name enables effective multi-query planning in seconds with three core components (Figure~\ref{fig:overview}):

\begin{denseitemize} 

\item \emph{Search Optimizer (\S\ref{sec:search-optimizer}):} 
Minimizes search steps (\ie, plans proposed and profiled) to find SLO-compliant plans. 
It optimizes the configuration ($c$), placement ($p$), and resource allocation ($r$) of operators, enabling flexible downstream scheduling.

\item \emph{SLO Profiler (\S\ref{sec:profiler}):} 
Efficiently profiles each plan's performance (e.g., accuracy) and identifies a set of Pareto-optimal plans to guide the search optimizer.

\item \emph{Multi-query Scheduler (\S\ref{sec:multi-query}):} 
Given a list of candidate plans per query, it selects query-plan combinations to maximize goodput.
\end{denseitemize}

\bipar{Workflow.}
\name employs a two-stage workflow for high efficiency and decision quality (Figure~\ref{fig:overview}).
\circled{1} When a user submits a new query, the Search Optimizer initiates an iterative single-query search process to generate (propose) candidate plans. 
\circled{2} The SLO Profiler evaluates the candidate's performance (\eg, accuracy and latency), filtering out infeasible plans that fail to meet SLO requirements. 
\circled{3} The Search Optimizer refines feasible candidates by optimizing their Pareto-optimal resource cost. This \circled{2}-\circled{3} repeats until a predefined termination condition is met (\eg, response-time limit) or the search space is exhausted.
\circled{4} The resulting candidate set is passed to the Multi-query Scheduler, which selects the best combination of query plans across all queries to maximize service goodput while avoiding SLO violations due to resource contention. The finalized deployment plan is then executed by the cluster resource orchestrator.
If runtime dynamics are detected, the resource orchestrator will trigger \name to start a replanning phase. 

%% file: sections/design.tex
\section{\name Design}
\label{sec:design}

\name decomposes the intractable multi-query planning into a two-stage planning process for scalability and adaptability. This design isolates performance fluctuations within a single query, avoiding costly system-wide replanning, while, for the first time, enabling real-time query planning for various deployment scenarios using a unified planner. However, such decomposition is non-trivial, introducing three unique challenges in balancing planning efficiency and decision quality:

\begin{denseitemize}
    \item \emph{Search Efficiency:} Each query may yield thousands of potential plan candidates with varying performance. How to minimize search steps to identify SLO-compliant candidates, improving responsiveness and planning costs (due to profiling) (\S\ref{sec:search-optimizer})?
    \item \emph{Profiling Efficiency:} Each proposed plan requires costly performance profiling (\eg, measuring final model accuracy using GPUs~\cite{zhang2024vulcan}). How to reduce the per-candidate profiling time and the overall monetary cost due to high query volumes (\S\ref{sec:profiler})?
    \item \emph{Resource Efficiency:} Independent query planning overlooks cross-query resource contention, potentially overloading cloud tiers while underutilizing edge resources. How to maintain high decision quality due to decomposition to maximize goodput (\S\ref{sec:multi-query})?
\end{denseitemize}

\begin{figure}[t]
    \centering
    \includegraphics[width=1.0\linewidth]{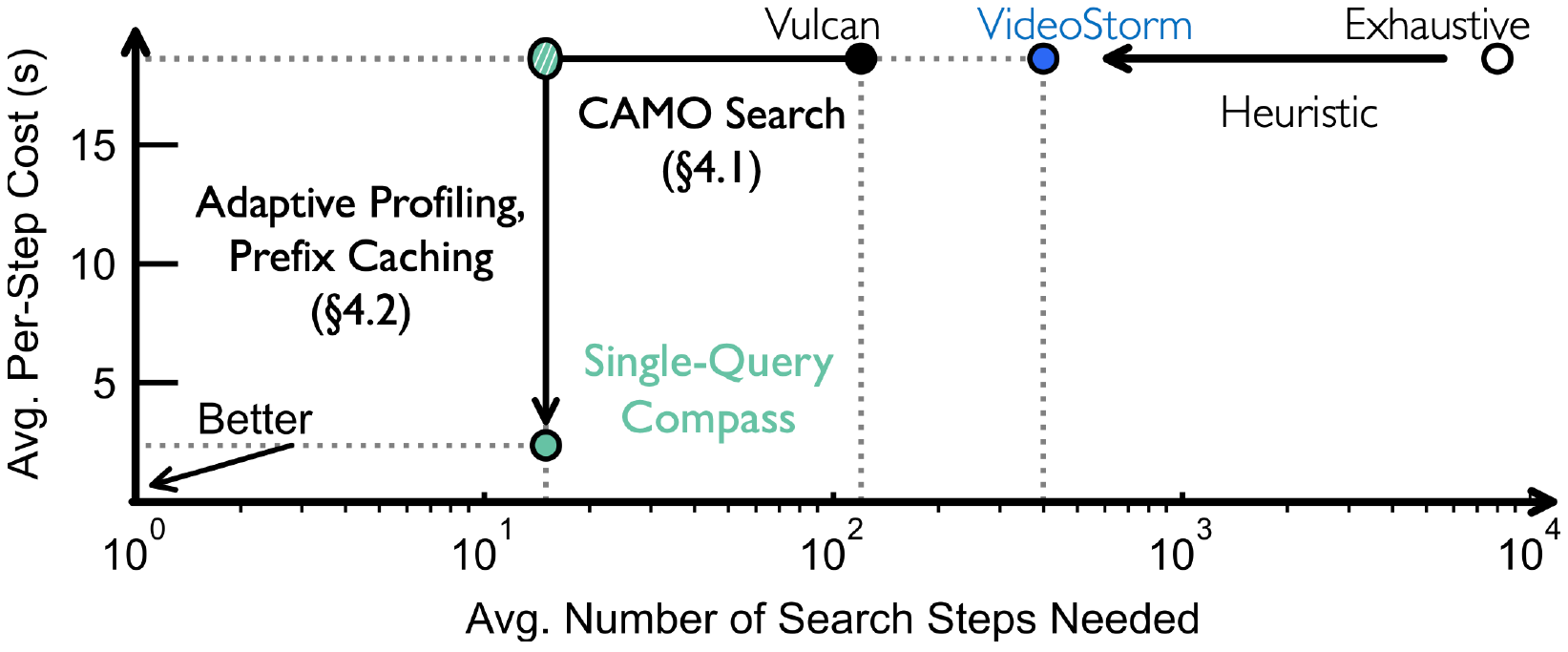}
    \caption{\name optimizes single-query search by reducing the number of search steps (\S\ref{sec:search-optimizer}) and per search step cost (\S\ref{sec:profiler}).}
    \label{fig:single-goal}
\end{figure}

Next, as shown in Figure~\ref{fig:single-goal}, we start by describing how \name enables real-time query planning by simultaneously reducing the per-step cost and the total number of steps required.

\subsection{Search Optimizer: Leverage  Similarities}
\label{sec:search-optimizer}

Even for a single query, identifying the optimal plan that satisfies latency and accuracy SLOs while minimizing deployment costs is challenging due to the presence of thousands of candidate plans. Worse, per-step search costs vary widely due to heterogeneous profiling costs; for example, searching (\ie, proposing and profiling) a plan involving LLaMA3-70B costs 9$\times$ more GPU time than its LLaMA3-8B counterpart. Consequently, minimizing the number of search steps is critical for both single- and multi-query planning.

To this end, \name employs a novel cost-aware search optimizer built on two key insights. 
First, reducing resource allocations to operators increases latency but does not affect pipeline accuracy. 
\revision{By \emph{tentatively} assuming resource over-provisioning in the plan (e.g., dedicating a cloud GPU to the query), \name defer fine-grained resource allocation to the multi-query scheduling stage (\S\ref{sec:multi-query}) after resource trimming (\S\ref{sec:profiler})} \footnote{Profiling over-provisioned plans adds no extra overhead, as the same data is processed to assess accuracy. So the amount of profiling execution is the same.}
\revision{This preserves optimality, as over-provisioned plans form a superset that contains the global optimum:  (1) they include plans meeting accuracy SLOs regardless of high resource costs; and (2) because resources are overcommitted, these plans represent the best possible latency of their counterparts with fewer resource allocations (e.g., when sharing the resource). 
Plans that still fail to meet latency or accuracy are naturally pruned without compromising decision quality. We defer the resource trimming design to the SLO Profiler (\S\ref{sec:profiler}), which finds the Pareto-optimal counterparts of these over-provisioned plans. We empirically validate this close-to-optimal performance across various settings (\S\ref{eval:ablation-study}).}

Second, instead of exploring plans independently, we can exploit their latent similarities to prioritize candidates with higher potential. However, plans are heterogeneous, comprising diverse operators and configurations, making explicit similarity metrics infeasible.

\bipar{Leveraging Intra-query Plan Similarity.}

\begin{figure}[t]
    \centering
    \includegraphics[width=0.99\linewidth]{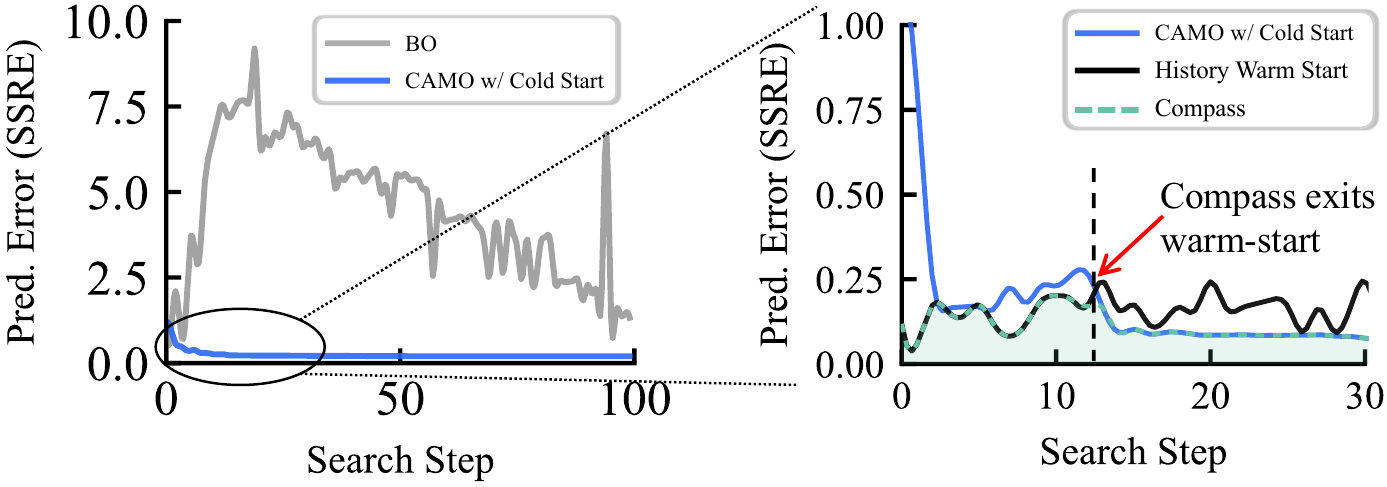}

    \caption{Prediction accuracy of $f_l$. Left: CAMO achieves accurate latency predictions, whereas standard BO shows significant errors. Right: \name leverages a committee of past models for warm start, transitioning to its own CAMO as it matures.}
    \label{fig:bo-combined}
\end{figure}

\name introduces a Cost-Aware Multi-Objective (CAMO) search optimizer that exploits the latent similarity across a query's candidate plans. CAMO builds on Bayesian Optimization (BO)'s core strength---learning from sparse, noisy feedback without requiring explicit performance models---but fundamentally departs from BO's traditional formulation. Classic BO assumes a single optimization objective and uniform evaluation costs. In contrast, compound AI pipelines exhibit inherently multi-dimensional objectives (accuracy, latency, and cost) and highly heterogeneous search costs, where profiling different plans can differ by orders of magnitude. As shown in Figure~\ref{fig:bo-combined} (left) and further validated in Section~\ref{eval:e2e}, these violations cause standard BO to mis-rank candidates, over-explore expensive regions, and ultimately become ineffective for compound AI serving.

To address these limitations, CAMO maintains two independent surrogate BO models: $f_a$ predicts the accuracy of a plan $p$, and $f_l$ predicts its latency. \name proposes new plans that maximize the acquisition function (utility) $U$:
\begin{equation*}
    U(p) = \textbf{Pr}[f_a(p)\geq A_{slo}]\cdot\,\textbf{Pr}[f_l(p) \leq L_{slo}] \times \frac{1}{\textbf{C}(f_l(p))}
\end{equation*}

where $\textbf{Pr}[\cdot]$ is the probability derived from the surrogate predictions. 
\revision{
$\textbf{C}(f_l(p))$ is the profiling cost function, which is essentially the latency to run the inference, so we can estimate the cost using predicted latency from our CAMO formulation.
}

\SetNlSty{large}{}{:}
\SetKwProg{Fn}{Function}{:}{}

\begin{algorithm}[t!]
    \small
    \caption{Compass Single-Query Search. \label{alg:single-query}}
    \SetKwFunction{CompassSingleSearch}{\textrm{SingleQuerySearch}}
    \SetKwFunction{ParetoOptimize}{\textrm{ParetoOptimize}}
    \SetKwFunction{Propose}{\textrm{Propose}}
    \SetKwFunction{Profile}{\textrm{Profile\_SLO}}
    \SetKwFunction{Update}{\textrm{CAMO\_Update}}
    \BlankLine

    \Fn{\CompassSingleSearch{$A_{slo}$, $L_{slo}$, dataset $\mathcal{D}$}} {
        
        $plans \gets []$ \\
        \While {$current\_time$ < $response\_time\_requirement$ \label{alg:search-start}} {
            \tcc{CAMO models of similar history queries vote for a new proposal if the new CAMO is immature.}
            $plan \gets  \Propose(C, A_{slo}, L_{slo}, f_A, f_L)$ \\
            \If{\Profile{plan, $\mathcal{D}$, $A_{slo}$, $f_A$, $f_L$}} {
                \tcc{If the plan is SLO-compliant, then generate its Pareto-optimal variants.}
                
                $plans \mathrel{+}= \ParetoOptimize(plan, L_{slo})$ 
            }
            \tcc{Update CAMO model with profiling feedback.}
            \Update($plans$, $f_A$, $f_L$)
        }

        \Return $plans$
        \label{alg:search-end}
    }

    \vspace{2mm}

    \Fn{\Propose{$C_r$, $A_{slo}$, $L_{slo}$, $f_A$, $f_L$}} {
        \tcc{When the CAMO model of that query matures, it proposes the plan with the highest utility.}
        \If{pred. error of $C_r$ is lower than committe's \label{alg:propose-start}} {
            $c_0 \gets \text{argmax}(C_r, x:\text{utility}(x, A_{slo}, L_{slo}, f_A, f_L))$ \\
        }\label{alg:cbo-propose}
        \Else {
            $c_0 \gets committe\_propose(C_r, A_{slo}, L_{slo})$ \\
        }
        \Return $c_0$
        \label{alg:propose-end}
    }

    \vspace{2mm}

    \Fn{\ParetoOptimize{c, p, $L_{slo}$}} {
     \tcc{Multi-dimensional binary search to tighten resource usage until the latency SLO is violated.}
        $r \gets over\_provisioned(c)$ \label{alg:tree-start}\\
        $candidates \gets multi\_dim\_binary\_search(c, p, r, L_{slo})$ \\
        \Return candidates
        \label{alg:tree-end}
    }

    \vspace{2mm}

    \Fn{\Update{plans, $f_A$, $f_L$}} {
        $f_A, f_L \gets fit\_new\_points(plans.acc, plans.lat)$ \\
        
        $update\_committe\_similarity(plans.acc, plans.lat)$
    }


\end{algorithm}


\revision{
This design provides three key advantages. First, by modeling accuracy and latency as separate objectives, CAMO evaluates each SLO's feasibility directly rather than collapsing them into a single scalar that obscures user requirements. 
Second, by explicitly incorporating profiling cost, CAMO not only prioritizes high-potential candidates but also prefers those that can be evaluated efficiently (benefit-to-cost ratio). This enables a more principled exploration-exploitation trade-off, while avoiding spending excessive time  on marginally beneficial but costly configurations for responsiveness. 
}
Third, decomposing objectives at the SLO level enables model reuse: CAMO can apply the latency surrogate learned from one query and the accuracy surrogate from another, as we will introduce shortly.

As shown in Algorithm~\ref{alg:single-query}, 
at each search step, the plan (\ie, the configuration and placement pair of operators) with the highest utility is proposed and sent to the SLO Profiler (Line~\ref{alg:search-start}-\ref{alg:search-end}). The profiling results serve as feedback to update CAMO's surrogate models. 
Figure~\ref{fig:bo-combined} (left) illustrates the effectiveness of our CAMO design over the standard BO solution. Note that our CAMO optimizer is pretty lightweight, requiring 40 ms per proposing step (\S\ref{eval:perf-breakdown}).

\bipar{Leveraging Inter-query Plan Similarity.}

\revision{Despite the rich heterogeneity in system conditions, input data, and QoE requirements, as well as the complex interplay among them, different queries may share latent structural patterns. Exploiting such inter-query similarity has been largely overlooked, yet it can dramatically reduce search time.
However, identifying similar past queries is challenging. Application-level metadata is frequently unavailable; even when present, it is impractical to encode diverse features for various queries (\eg, multimodal inputs, operator graphs, or model-configuration knobs). Compass leverages historical CAMO models, which have already learned these latent structures, to warm-start the search process for new, similar queries.}



\revision{\name uses CAMO's latent representation of candidate plans to define similarity by prediction error, \ie, the gap between a past surrogate model's prediction and the current query's real profiled performance. This requires no additional query information and lets CAMO reuse latently similar accuracy and latency models from different prior queries.}

Algorithm~\ref{alg:single-query} realizes this idea through a \emph{consensus-guided warm-start}. Instead of selecting a single most similar query, which is fragile in the presence of noise, dynamics, or partial alignment, \name assembles a committee of past surrogate models whose errors are lower than the current CAMO optimizer. Each committee member votes for the next candidate plan to propose, with a weight proportional to its similarity to the new uery (Lines~\ref{alg:propose-start}--\ref{alg:propose-end}), effectively pooling cross-query inductive biases. Then, the selected plan is profiled, and similarity scores are updated based on prediction errors, allowing the committee to reorganize itself on-the-fly.

A novel emergent behavior of our consensus-guided design is its self-terminating transfer. 
As shown in Figure~\ref{fig:bo-combined} (right), once the new CAMO optimizer matures---\ie, when its prediction error is lower than the committee---\name automatically exits the warm-up phase, allowing the search to focus on the new query's unique context (Line~\ref{alg:cbo-propose}). 
As CAMO is lightweight in proposing plans, we further empirically validate that the warm-start design adds little overhead while delivering substantial speedups, even with only a dozen prior queries in history (\S\ref{eval:ablation-study}).

\subsection{SLO Profiler: Ensure Correctness Efficiently}
\label{sec:profiler}

For every candidate plan proposed by the Search Optimizer, verifying its SLO compliance through profiling is essential for steering the search process. 
Prior ML serving advances~\cite{wu2022jellybean} have streamlined profiling via two well-established insights. 
\revision{First, a plan's accuracy is determined by its operators' configuration, not by placement or resource allocation, and can therefore be measured efficiently in the cloud (e.g., using proxy inputs or replayed recent inputs~\cite{zhang2017videostorm}).  
Second, pipeline latency can be decomposed into operator runtimes and their communication costs. Since operator outputs are invariant to placement and resource settings, existing deployments~\cite{zhang2024vulcan, jiang2018chameleon, walle-osdi22} reuse operator-level latency from accuracy profiling and estimate end-to-end latency by applying resource- and input-dependent offsets, such as adjusting data transfer time based on network bandwidth. }
Recent work~\cite{wang-nsdi2025simai,kossmann2024cascadeserveunlockingmodelcascades,rashidi2020astrasim,bang2023vtrain} validates the effectiveness of this approach. So both accuracy and latency profiling can be consolidated into the accuracy profiling phase.

However, as ML models and pipelines grow more complex, accuracy profiling has become a major bottleneck. Profiling requires running GPU inference over many data samples: profiling \emph{a single} speech recognition plan can take 5-10 minutes on an A100 GPU, driving the total profiling cost to hundreds of dollars when scaled to thousands of plan candidates for a single query (\S\ref{eval:perf-breakdown}). 
Unfortunately, accuracy is highly input-dependent, varying across input distribution and user context (\S\ref{sec:background}). Even minor configuration changes (\eg, switching video resolution) can alter the input data, propagate through the pipeline, and impact the overall accuracy.

\begin{figure}[t]
    \centering
    \subfigure[Shaded area: 1-99\% range of estimated accuracy over 100 profiling runs for a plan. Solid line: one trial run. Red-dotted line shows overestimation from insufficient profiling. Verdict depends on the gap between ground truth and the SLO. \label{fig:adaptive-profiling}]{\includegraphics[width=0.48\linewidth]{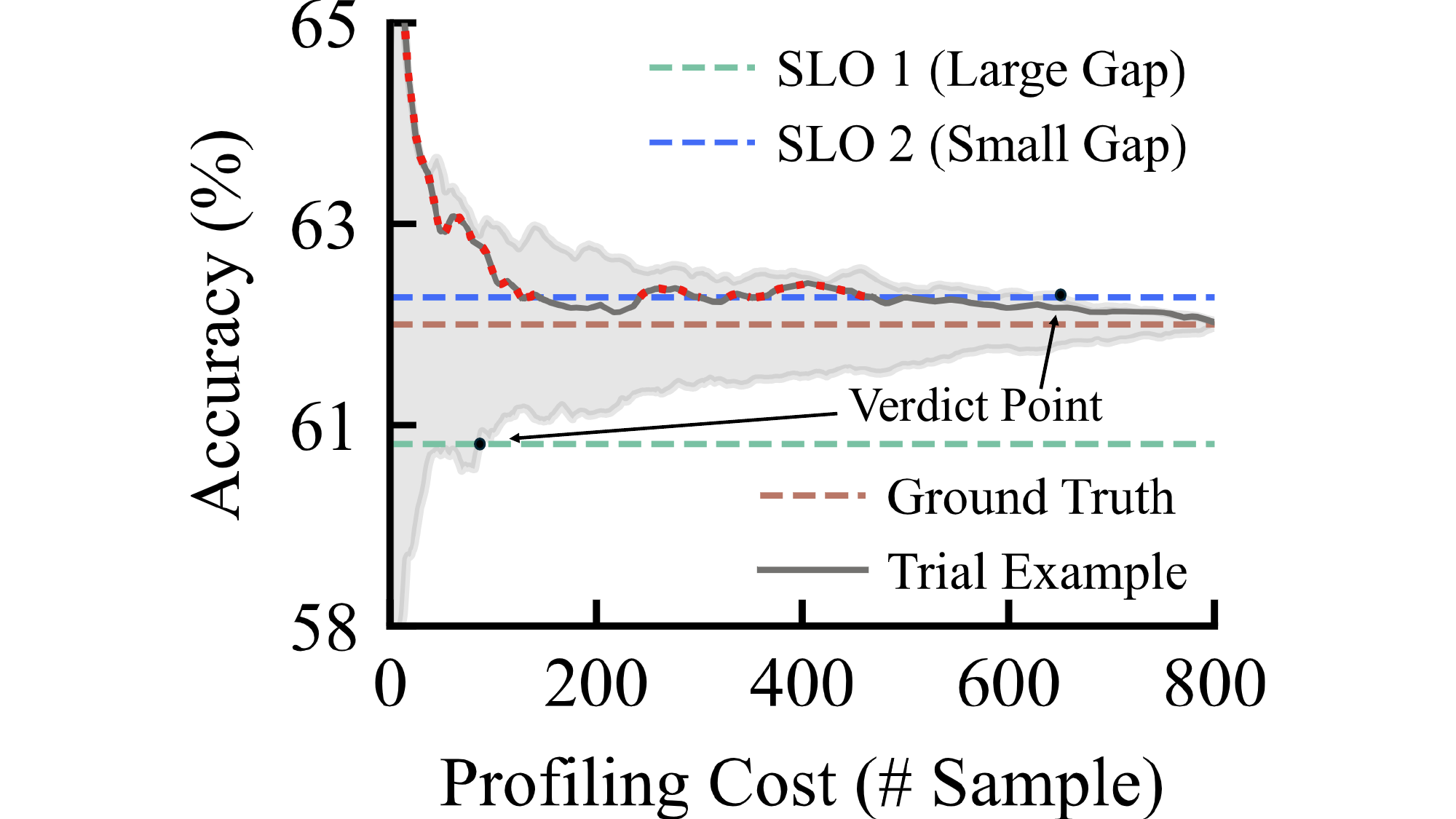}
        }
        \hfill
\subfigure[Adaptive profiling ensures the same correctness level (e.g., 99\%) while adjusting the number of samples per plan, significantly reducing the amount of profiling compared to fixed profiling designs (e.g., using a uniform Y=99\% cutoff). 
\label{fig:ttest-lorenz}]
{\includegraphics[width=0.48\linewidth]{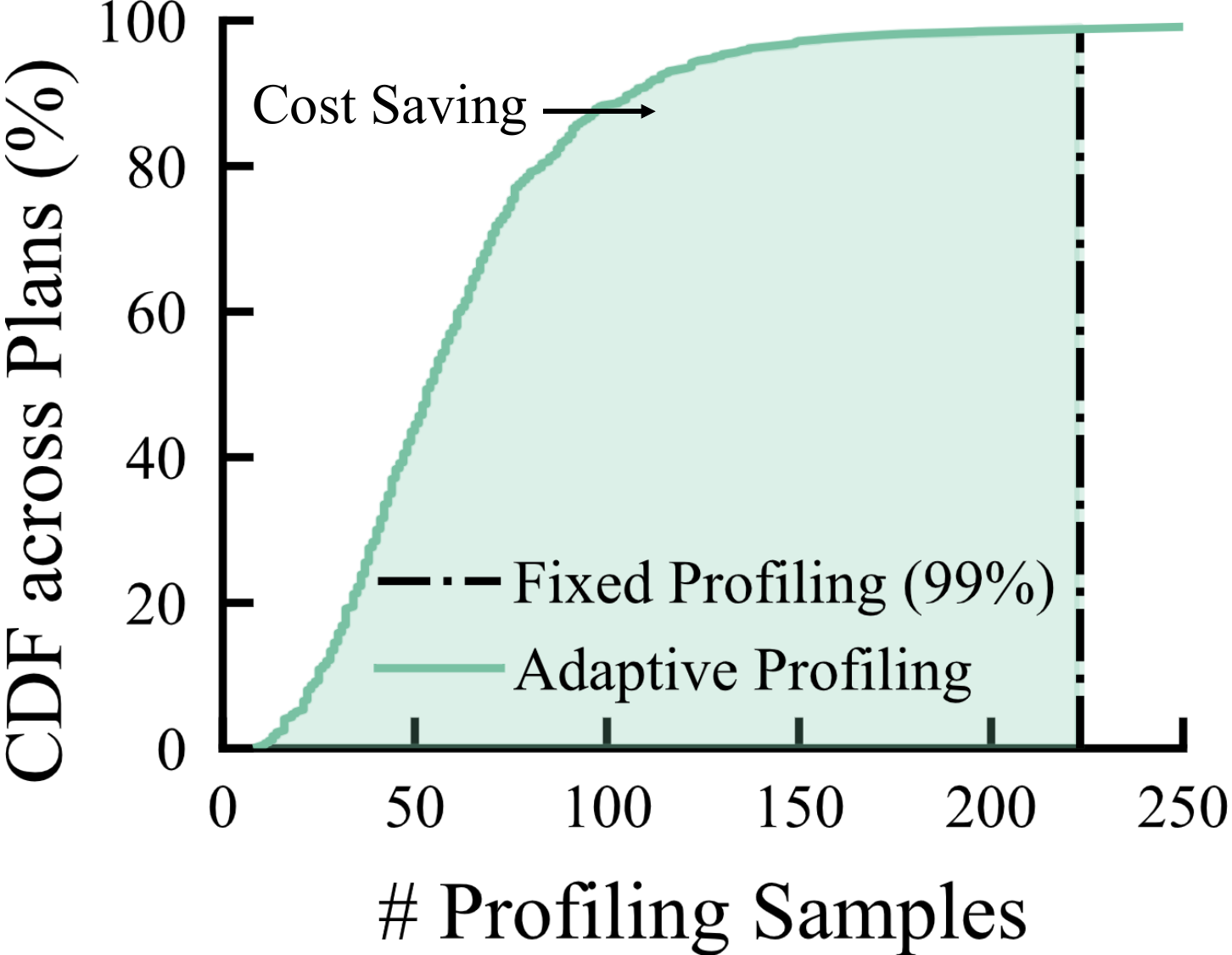}}
    \caption{Adaptive profiling ensures correctness and efficiency.}
\end{figure}

\name reduces this increasingly dominant cost and enables real-time planning with two complementary designs:
(i) adaptive profiling, which minimizes required profiling data size while preserving correctness; and
(ii) dependency-aware caching, which eliminates redundant profiling by reusing overlapping operator profiles.

\bipar{Adaptive Profiling for Correctness with Minimal Work.}
Profiling accuracy using the full dataset $\mathcal{D}$ yields ground truth but is prohibitively expensive. Prior efforts~\cite{wu2022jellybean, zhang2017videostorm, focus-osdi2018} typically rely on a fixed, manually specified number of profiling samples (e.g., 500 video frames). As shown in Figure~\ref{fig:adaptive-profiling}, this ad-hoc design creates a tradeoff between profiling cost and correctness: (1) too few samples jeopardize correctness, mislabeling invalid plans as SLO-compliant and ultimately causing SLO violations and misleading the search optimizer; but (2) too many samples waste GPUs and slow down search steps. In Figure~\ref{fig:adaptive-profiling}, stopping in the red dotted part of the profiling trial leads to an incorrect verdict for SLO target 2, while stopping after the verdict point is a waste. 

\name resolves this tradeoff with insights that profiling only needs to determine whether a plan meets the SLO (\ie, $\mathbb{I}(\text{acc} \geq \text{acc}_{\text{slo}})$), and the required sample size depends on the gap between the plan's true accuracy and the SLO target.  
As shown in Figure~\ref{fig:adaptive-profiling}, different SLO targets for the same plan require different amounts of profiling data (thus time): a large gap allows an early verdict with few samples (SLO 1), while a small gap demands many more samples for a confident decision (SLO 2) due to sample variance.  

\name proposes adaptive profiling that terminates once the SLO verdict is statistically certain, minimizing overhead. Specifically, \name adopts the Student's t-test~\cite{t-test-nips}, continuously estimates confidence using the accuracy sample mean and its variance on the profiled data's distribution thus far, which reflects the likelihood that the plan satisfies the SLO threshold.
Profiling stops once confidence exceeds a target level (e.g., 99\%), with theoretical guarantees. As shown in Figure~\ref{fig:ttest-lorenz}, this adaptive profiling achieves both efficiency and strong correctness guarantees. 

\bipar{Dependency-shared Profiling.}
Candidate-plan search often revisits overlapping pipeline segments, introducing opportunities for incremental profiling. Specifically, changes to downstream operators do not affect upstream outputs. This structural dependency creates opportunities for incremental profiling, but existing advances~\cite{zhang2024vulcan, wu2022jellybean} treat accuracy profiling as a monolithic end-to-end procedure and therefore repeatedly profile identical segments.

\name introduces dependency-shared profiling, a profiling-time optimization that identifies and reuses shared execution across candidate plans. The SLO Profiler maintains operator-level dependency information and selectively caches intermediate outputs at strategic boundaries, especially for GPU-expensive stages. When a new plan is proposed, the profiler performs a fast prefix match against cached subgraphs and skips redundant execution for all reused upstream portions, profiling only the operators that change.

The cache involved in dependency-shared profiling is scoped to a single query's planning session and discarded immediately after, keeping memory overhead \underline{\emph{transient}} and small (e.g., about 60 MB throughout a code-generation query). Note that GPU computation, not caching I/O, is the primary bottleneck during profiling, and storage can be scaled more cost-effectively across machines than expensive GPUs (\S\ref{eval:perf-breakdown}); thus dependency-shared profiling delivers savings without introducing new bottlenecks.

\bipar{Resource Trimming: Identify Resource-Pareto Plans.}
\revision{As the Search Optimizer tentatively over-commits resources, the resulting plans may not lie on the cost-optimal Pareto frontier preferred by service providers. 
To address this, for each identified SLO-compliant plan, the SLO Profiler refines the plan's resource costs via a greedy reduction process (Line~\ref{alg:tree-start}--Line~\ref{alg:tree-end} in Algorithm~\ref{alg:single-query}). Based on the monotone property that reducing resources increases latency without affecting accuracy, \name performs a multi-dimensional binary search to iteratively trim operator-level resources (e.g., reducing GPU share from 100\% to 50\% supported by MPS~\cite{nvidiamps}).
The trimming procedure halts once further reductions would violate latency constraints. 
This process produces a rich set of low-cost, SLO-compliant plans along the Pareto frontier, which supply the multi-pipeline planner for maximizing goodput.}

\subsection{Multi-Query Planner: Global Scheduling}
\label{sec:multi-query}

So far, \name efficiently identifies cost-effective plans for individual queries. However, when multiple queries share the same infrastructure, naively deploying each query's locally optimal plan can lead to severe global inefficiencies (\S\ref{sec:opportunity}): some machines or tiers become overloaded while others remain underutilized, ultimately reducing overall SLO goodput. To address this, \name extends its single-query planning to multi-query global scheduling, targeting two practical deployment objectives: (i) maximize SLO service goodput under resource-constrained environments (e.g., private clusters), and (ii) minimize total deployment costs when resources can scale elastically (e.g., public clouds), efficiently serving all queries.

\bipar{Maximizing SLO Goodput.} 
In resource-constrained deployment spaces, it is often infeasible to serve all queries concurrently, so maximizing SLO goodput becomes critical. 
To preserve high-quality planning decisions due to decomposition, \name goes beyond selecting the locally optimal plan for each query (\ie, the lowest-cost SLO-compliant plan). 
Instead, it leverages the full set of Pareto-optimal plans identified by the single-query planner. These candidates span different resource footprints and infrastructure tiers while meeting SLO requirements.

However, deciding query-plan combinations to maximize goodput is notoriously difficult, resembling a \emph{graph bipartite matching} problem between the query set and plan set, combined with a \emph{multi-dimensional maximum cardinality bin packing} problem to allocate operators into machines. In fact, even a relaxed version---where each query is restricted to its single best plan---is still NP-hard~\cite{chekuri2004multidimensional-apxhard}.  
One potential formulation is as an integer linear program (ILP): deciding which queries to serve, selecting plans from available options, and assigning them to machines, subject to per-tier resource constraints, with the objective of maximizing total SLO goodput. We present the full ILP in Appendix~\ref{appendix:mp-proof}. 
Unfortunately, for $N$ queries, each with $P$ feasible plans, across $T$ tiers and $M$ machines per tier, this formulation yields $O(N \cdot P \cdot T \cdot M)$ binary variables, intractable for large-scale deployments.

To scale, \name leverages a greedy heuristic. Each plan $i$ has an \emph{aggregate} cross-tier and machine resource demand $cr_i$ and an associated SLO benefit $w_i$. \name ranks all plans across queries based on their benefit-to-resource ratio (\ie, $w_i$/$cr_i$). 
The planner then iteratively allocates resources to the highest-ranked plans, adding a plan to the allocation if its corresponding query has not yet been scheduled. This process continues until available resources are exhausted.

\bipar{Minimizing Total Deployment Costs.} 
Unlike the setup with limited resources, service providers may rely on public clouds with elastic resource scaling to serve all queries, where minimizing deployment costs becomes the key. To address so, our planner extends our previous solution by modifying the sorting criteria. Instead of prioritizing based on the benefit-to-resource ratio, plans are ranked and selected based on their SLO benefit-to-monetary cost ratio.

Our greedy heuristic is flexible and can incorporate various practical considerations. For example, service providers can dynamically assign higher weights to queries as their pending time increases to prevent starvation and ensure fairness. Additional costs (e.g., migration overheads) can be injected into the cost term too. 
In fact, our multi-query planner offers a theoretical guarantee:
\begin{highlighttheorem}
\begin{theorem}
\label{theorem:mp}
Let $A(t)$ and $C(t)$ denote the goodput and deployment cost achieved by our greedy planner at time~$t$. $A_{\mathrm{opt}}(t)$ and $C_{\mathrm{opt}}(t)$ denote the optimal values. Then:
\begin{denseenum}
    \item In cloud-only setups, $A(t) \geq \frac{1}{2} A_{opt}(t)$; 
    \item Let $d$ be the number of tiers, $C(t)\leq \frac{11}{9}\times C_{opt}(t) + O(d)$.
\end{denseenum}
\end{theorem}
\end{highlighttheorem}
\revision{We leave the detailed proof in  Appendix~\ref{appendix:mp-proof}. 
It shows that the deployment cost from our greedy planner is within a competitive ratio of $\frac{11}{9}$ to the optimal. Indeed, our empirical evaluations show \name achieves comparable performance with oracle method (\ie, ILP with unlimited search time), while preserving scalability, even in large-scale deployments (\S\ref{eval:e2e} and \S\ref{eval:ablation-study}).}

\bipar{Online Adaptation.}
\revision{In real-world deployments, query input distributions and runtime environments may change over time (\eg, lighting changes or network fluctuations in GenAI chat~\cite{chatgpt-videochat}), risking SLO violations. Despite various potential dynamic sources (adaptation needs) such as multi-tier resource changes and data distribution shifts, online adaptation overhead primarily boils down to three sources: (i) single-query search adaptation, (ii) multi-query scheduling adaptation, and (iii) execution-level migration. 
Building on existing infrastructure support for pipeline migration~\cite{gcp-iot, zhang2024vulcan, k8s}, the core challenge becomes how to efficiently replan in the wild.
When triggered by the service monitor~\cite{gcp-iot} or user intervention, \name warm-starts the previously learned CAMO model, reusing prior knowledge to quickly correct stale estimations and adapt to the new environment. Our component-wise and end-to-end evaluations show that \name's robust responsiveness in seconds (\S\ref{eval:ablation-study}).
}

%% file: sections/implementation.tex
\section{Implementation}
\label{sec:implementation}

We implemented a \name prototype in roughly 2,800 lines of Python, fully compatible with Kubernetes~\cite{k8s} and LangChain~\cite{langchain}, and supporting existing ML stacks with minimal API changes.


\begin{table*}[ht]
  \centering
  \small
  \resizebox{0.99\linewidth}{!}{
  \begin{tabular}{l|l|c|l}
      \hline
      \textbf{Task} & \textbf{Model Families} & \textbf{Search Space} & \textbf{Description}   \\
      \hline
      Live Agent Chat (LVC) & InternVL2~\cite{chen2024internvl}, Gemma3~\cite{gemma3-technical-report} & $\sim$ 7K & Video conversation with multimodal models.\\
      Agentic Code Generation (ACG) & Qwen3~\cite{qwen2-report}, Llama3~\cite{dubey2024llama}, DeepSeek~\cite{deepseek-coder-technical-report} & $\sim$ 1K & Code generation via two agents. \\
      Dense Video Captioning (DVC) & Whisper~\cite{radford2023robust}, T5~\cite{raffel2020exploring}, Vid2Seq~\cite{yang2023vid2seq} & $\sim$ 23K & Segments and geneartes captions for a video. \\
      Visual Tracking (VT) & YOLO~\cite{redmon2016you}, ResNet~\cite{he2015resnet} & $\sim$ 6.5K & Object detection and tracking from edge camera.\\
      Speech Recognition (SR) & Wave2Vec~\cite{baevski2020wav2vec}, Hubert~\cite{hsu2021hubert} & $\sim$ 6.4K & Converts live human speech into text. \\
      \hline
  \end{tabular}
  }
  \caption{\name supports representative, real-world compound AI applications. Task and configuration details are available in Appendix~\ref{appendix:setup}.}
  \label{tab:models}
\end{table*}

\begin{table*}[ht]
  \centering
  \small
  \resizebox{0.99\linewidth}{!}{
  \begin{tabular}{l|l|l|l|l}
      \hline
      \textbf{Tier} & \textbf{Specification} &\textbf{GPU} & \textbf{CPU, Memory and In-cluster Network} & \textbf{Price}  \\
      \hline
      Cloud &  a2-ultragpu-4g & 4$\times$ Nvidia A100 (80GB HBM) w/ NVLink & 48 vCPU (Intel Cascade Lake), 680GB DRAM, 50 Gbps & 20.3\$/h  \\
      Near-Cloud &  n1+v100*4 & 4$\times$ Nvidia V100 (16GB HBM) & 32 vCPU (Intel Haswell),  160GB DRAM, 48 Gbps & 8.0\$/h  \\
      Near-Edge &  n1+t4 & 1$\times$ Nvidia T4 (16GB) & 8 vCPU (Intel Haswell),  40GB DRAM, 16 Gbps  & 1.3\$/h\\
      Edge &  N/A & Commdity GPUs (i.e. RTX 1060) and SoCs & Varies & Free \\ 
    \hline
  \end{tabular}
  }
  \caption{Our evaluation uses realistic multi-tier infrastructure, including 16 A100s and 16 V100s across tiers. Details are provided in Appendix~\ref{appendix:setup}.}
  \label{tab:cluster}
\end{table*}

\bipar{Interface, Runtime and Performance Isolation.}
\revision{\name exposes simple APIs for specifying SLOs (Appendix~\ref{app:interface}), covering the majority of compound AI workflows. Our current implementation uses Kubernetes to manage the cluster and containers for each operator. For the underlying inference engine of LLM workflows, we use vLLM~\cite{vllm-sosp23}. We leverage Linux cgroups and virtualization technologies like NVIDIA MPS and MIG and AMD MxGPU to enforce performance isolation, thus avoiding performance degradation from resource contention. Our implementations show only about 7\% performance interference with temporal GPU sharing. Note that  to further mitigate potential contention effects, we enable conservative resource partitioning during profiling. Rather than relying on fine-grained fractional allocations, we restrict GPU slices to coarse levels (e.g., 25\%, 50\%, 75\%, 100\%).}
Our Kubernetes implementation supports distributing operators across machines, connected via gRPC. 
Following existing advances~\cite{zhang2024vulcan, gcp-iot}, we maintain standby machines to minimize latency. 
\revision{We provide a detailed runtime latency breakdown, including state migration (\S\ref{eval:perf-breakdown}).}

\bipar{Fault Tolerance.}
\name periodically and asynchronously records profiled configurations to a checkpoint file.
In case of failure, \name can resume execution from the last saved state, reducing redundant profiling efforts.
For backend fault tolerance, we rely on cluster management tools such as Kubernetes to handle failures.






%% file: sections/evaluation.tex
\section{Evaluation}
\label{sec:eval}

We evaluate \name on five representative compound AI applications  (\eg, agentic video chat) across realistic deployment spaces (e.g., cloud-native and edge-cloud infrastructures), showing:

\begin{denseitemize}
    \item \name achieves over 2.4--5.1$\times$ better service goodput, and slashes deployment costs by 3.8--4.5$\times$ (\S\ref{eval:e2e}).
    
    \item \name accelerates single-query planning by 4.2--10.5$\times$, achieving second-level responsiveness (\S\ref{eval:e2e}).
    
    \item For multiple-query planning, \name effectively scales to thousands of queries while achieving \emph{close-to-optimal} global planning performance (\S\ref{eval:e2e}-\S\ref{eval:perf-breakdown}).
    
    \item \name improves performance across varied settings and reacts to dynamics in seconds (\S\ref{eval:ablation-study}). 

\end{denseitemize}

\subsection{Methodology}
\label{eval:setup}

\bipar{Realistic Infrastructure Setup.}
\name is designed to support large-scale deployments with numerous end users. Conducting experiments at full scale would be prohibitively expensive and challenging to reproduce. To balance realism and practicality, we adopt a mid-scale, multi-tier infrastructure that goes beyond prior evaluations~\cite{zhang2024vulcan, wu2022jellybean, slm-acl25}. As summarized in Table~\ref{tab:cluster}, our setup consists of four tiers involving 16 A100s and 16 V100s:
(i) a cloud tier with high-end machines, each equipped with 4 A100 GPUs on Google Cloud Platform;
(ii) a near-cloud tier with mediocre machines hosting 4 V100 GPUs;
(iii) a near-edge tier containing 1 T4 GPU; and
(iv) local edge devices, including heterogeneous commodity GPUs (e.g., RTX 1060) and smartphones with SoCs.
This heterogeneous deployment allows us to realistically emulate multi-tier workloads.


By default, we report performance on the medium-scale deployment spanning cloud, near-cloud, and edge tiers. The medium setup consists of 4 cloud machines, 4 near-cloud machines (Total 16 A100 GPUs + 16 V100 GPUs). We further consider large-scale scenarios using traces collected from above practical deployment (e.g., plan latency and accuracy), simulating 32 cloud machines, and 32 near-cloud machines (Total 128 A100 GPUs + 128 V100 GPUs). To show the generality of \name, we also evaluate a single-tier cloud-only setup and a four-tier medium-scale setup (\S\ref{eval:ablation-study}). 

\revision{
Due to space limit, additional experiment setup details (e.g., edge device detail and network bandwidth) and results can be found in Appendix~\ref{appendix:setup} and \ref{app:eval_results}, respectively.}



\begin{figure*}[t]
    \centering
        \includegraphics[width=.95\textwidth]{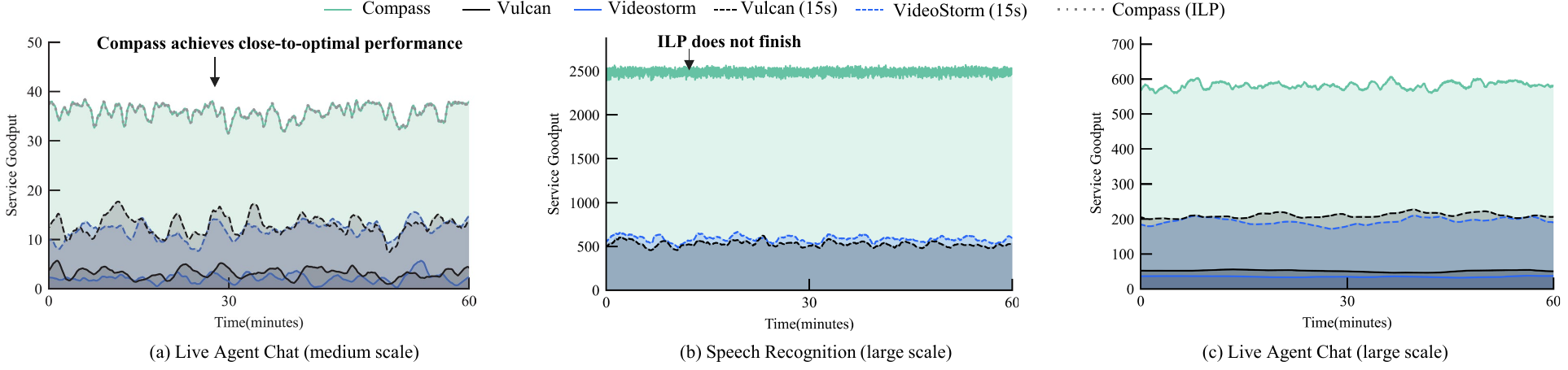}
    \caption{In resource-constrained deployments, \name improves service goodput across scales and applications, achieving performance close to the ILP optimal. It outperforms baselines that require 3$\times$ longer response times---15 seconds render real-time service impractical.}
    \label{fig:multi-pipeline-limited}
\end{figure*}

\bipar{Datasets, Queries, SLOs, and Loads.}
We evaluate \name using five popular compound AI applications with real-world workloads  (Table~\ref{tab:models}). Real-time applications, like live agent chat, visual tracking, and speech recognition, desire fast responsiveness, thus being \textit{latency-critical} in responsiveness. While agentic code generation and video captioning are batch applications targeting the most cost-effective plans, thus \textit{throughput-critical}. 
\revision{
We use realistic datasets with diverse input distributions and popular models (Appendix~\ref{appendix:setup}).}

We follow query workloads and SLOs inspired by recent advances in model-serving planning~\cite{zhang2023shepherd, infaas-atc21, zhang2017videostorm}, such as \textit{``track and imitate a user's body movements during push-ups and provide instant corrective feedback on muscle engagement''} in live agentic video chat. To construct realistic SLO requirements, we first identify the Pareto-optimal plans for each query via exhaustive search. We then relax and sample the accuracy and latency performance of plans on the Pareto frontier to generate diverse SLOs: by default, accuracy is set to 85\% and latency to 1.5$\times$ the sampled points. Uniform sampling of these Pareto-optimal plans ensures coverage of a broad search space. We further evaluate performance under varying SLO settings (\S\ref{eval:ablation-study}).

User query arrivals follow Microsoft's ML-serving traces~\cite{dynamollm-isca24}, scaled proportionally to our setup's total resource capacity. More specific, to simulate realistic system load, we use a Pareto-optimal plan to control query arrival rates. We define the cluster capacity as the maximum number of Pareto-optimal plans that can be served concurrently. System load \textbf{1} corresponds to an arrival rate matching this capacity, ensuring the cluster is fully utilized without overload. In practice, baselines cannot achieve Pareto-optimal plans, resulting in slight overload, which provides a meaningful comparison point. We validate the impact of different system loads in Section~\ref{eval:ablation-study}.

\bipar{Baselines.}
State-of-the-art methods focus on either single-tier or single-query settings. We extend them into \emph{stronger baselines} that operate in multi-tier, multi-query environments:
\begin{denseitemize}
    \item \textit{Vulcan} \cite{zhang2024vulcan}: 
    A state-of-the-art query planner designed for cross-tier, single-query live ML analytics. It encodes operator placement and configuration search into an aggregate utility function.
    
    \item \textit{VideoStorm}: a cloud-only multi-query planner that uses hill climbing~\cite{zhang2017videostorm}. 

\end{denseitemize}

We augment Vulcan with our multi-query planner to support many queries, and extend VideoStorm to support cross-tier settings by incorporating placement search into its hill-climbing method, alongside our multi-pipeline planner.
Our results are consistent with prior reports of the baselines.



\bipar{Metrics.}
We aim to improve the following metrics:
\begin{denseitemize}
\item \emph{Low response time}: the time to find the first SLO-compliant query plan, ensuring service responsiveness;

\item \emph{Low profiling cost}: the monetary cost due to profiling; 

\item \emph{High SLO service goodput}: the number of queries meeting SLO requirements under limited resource capacities. 

\item \emph{Low deployment cost}: 
the deployment costs to sustain all queries (\eg, in public clouds with scaling capabilities). 

\end{denseitemize}

We report the mean values over five runs per experiment.

\subsection{End-to-End Performance}
\label{eval:e2e}

We evaluate \name in online multi-query settings. 
For \textit{latency-critical} queries, we use a 5-second planning deadline, matching live-service startup constraints~\cite{infaas-atc21, paraserve-arxiv25, gemini-videochat}. Baselines also get a relaxed 15-second timeout.
For \textit{throughput-critical} queries, we use a 10 A100-GPU-hour profiling budget ($\sim$\$37 in GCP) per query.
We later present ablation studies under different response time and profiling cost budgets (\S\ref{eval:ablation-study}).

\bipar{\name achieves SLO service goodput close to optimal.}
Figure~\ref{fig:multi-pipeline-limited} shows that, over a 60-minute service window, \name achieves 15--61$\times$ higher average service goodput compared to Vulcan and VideoStorm in both medium- and large-scale deployments. The improvements are consistent across tasks and sustained over time. Even when compared to Vulcan (15s) and VideoStorm (15s), \name improves average SLO goodput by 2.4--5.1$\times$, respectively. Note that 15 seconds have rendered many real-time services impractical~\cite{infaas-atc21}. 

Crucially, \name delivers performance comparable to ILP-based scheduling, showing that our greedy scheduler approximates the global optimum. Unlike ILP solvers, which do not scale to large deployments, \name's scheduling algorithm scales sub-linearly (see Appendix~\ref{app:eval_results}), enabling consistent, at-scale improvements. We later validate \name's comparable performance with an exhaustive search-based oracle (Figure~\ref{fig:abla-optimal-scalable}).

\begin{figure}[t]
\centering
\subfigure[Visual Tracking]{
\includegraphics[width=0.47\linewidth]{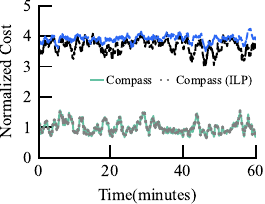}
}
\hfill
\subfigure[Speech Recognition]{
\includegraphics[width=0.47\linewidth]{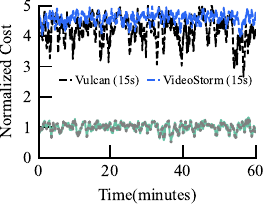}
}
\caption{In cost-constrained deployment, \name reduces resource monetary costs in supporting all queries.}
\label{fig:multi-pipeline-unlimited}
\end{figure}

\begin{figure}[t]
  \centering
    {
    \subfigure[Dense Video Captioning\label{fig:time2cost-dvc}]{\includegraphics[width=0.49\linewidth]{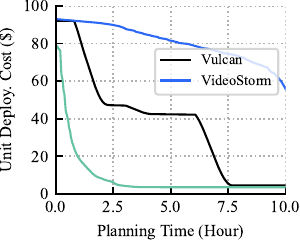}}
    \hfill
    \subfigure[Agentic Code Generation\label{fig:time2cost-acg}]{\includegraphics[width=0.49\linewidth]{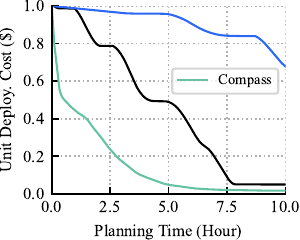}} 
    
  }
  \caption{\name reduces planning (profiling) costs for throughput-critical applications, in A100 GPU hours.}
  \label{fig:time-cost}
\end{figure}

\bipar{\name slashes deployment costs.}
We next evaluate the resource-abundant deployment scenario (\eg, utilizing cloud scaling services). Monetary cost is measured using Google Cloud Platform prices. For \textit{latency-critical} tasks, we report the real-time normalized monetary cost over 60 minutes; for \textit{throughput-critical} tasks, we report the unit (per hour) deployment cost of a single query pipeline.
Figure~\ref{fig:multi-pipeline-unlimited} shows that \name lowers the monetary cost for serving all queries by 3.8--4.5$\times$  compared to baselines. Figure~\ref{fig:time-cost} shows that \name finds a 2.0--6.0$\times$ cheaper deployment plan in time. 

\bipar{\name enables second-level responsiveness while reducing planning expense.}

\begin{figure}
    \centering
    \includegraphics[width=.95\linewidth]{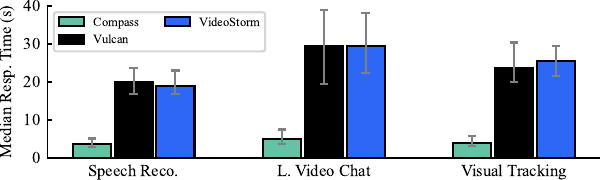}
    \caption{\name improves query responsiveness in single-query planning. Error bars show 25\% - 75\% percentile.}
    \label{fig:single-ttft}
\end{figure}

Figure~\ref{fig:single-ttft} shows \name identifies the first SLO-compliant plan in five seconds, 4.2--10.5$\times$ faster than Vulcan and 6.0--12.3$\times$ faster than VideoStorm. 
This fast responsiveness is important for user engagement in real-time services~\cite{dynamollm-isca24}. 
Moreover, \name spends 6$\times$ less time finding a comparable plan (Figure \ref{fig:time-cost}) for \textit{throughput-critical} queries, meaning substantial profiling cost savings considering high query volumes. 

\subsection{Performance Breakdown}
\label{eval:perf-breakdown}



\begin{figure}[t]
  \centering
    {
    \subfigure[End-to-end Perf. Breakdown. \label{fig:e2e-breakdown}]{\includegraphics[width=0.48\linewidth]{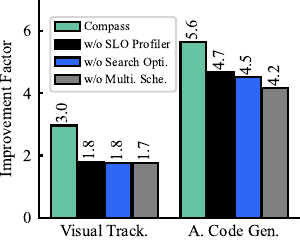}}
    \hfill
    \subfigure[Planning Runtime Breakdown.\label{fig:system-overhead}]{\includegraphics[width=0.48\linewidth]{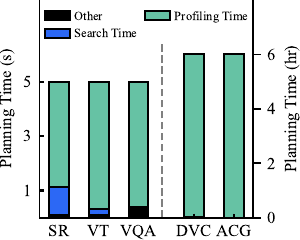}} 
  }
  \caption{Performance breakdown of \name.}
\end{figure}

\bipar{Breakdown by system components.}
We analyze the improvement factor of \name's each component: (i) \textit{Compass w/o SLO Profiler}: removes \textit{SLO Profiler}, using fixed sampling without prefix cache,
(ii) \textit{Compass w/o Search Optimizer}: removes history warm start and CAMO, using BO in Figure~\ref{fig:bo-combined} (left), 
and (iii) \textit{Compass w/o Multi-query Planner}: uses Yarn~\cite{yarn-socc13} scheduling (admits queries in a first-come-first-served fashion) for many queries, using their locally optimal plans. 
We use VideoStorm-based local planning as the 1$\times$ baseline, with the same planning time budgets.
Figure~\ref{fig:e2e-breakdown} shows each component of \name achieves 1.7--4.7$\times$ comparable improvements, contributing to 3.0--5.6$\times$ end-to-end improvements.

\bipar{Breakdown by planning runtime.} 
\revision{
We next decompose the single-query planning overhead, primarily consisting of three aspects: (i) \textit{Profiling Time}: Time spent on SLO profiler, profiling the plan's SLO performance on GPUs (\eg, model inference overhead); 
(ii) \textit{Search Time}: 
Time spent on the Search Optimizer, including CAMO updates and plan proposing; 
and (iii) \textit{Other}: 
Miscellaneous system overhead, such as setting up connections within the Kubernetes backend. This includes time to spin up pre-warmed containers.
Figure~\ref{fig:system-overhead} shows that for latency-critical tasks, profiling dominates the time budget, accounting for over 80\% of the total planning overhead. For throughput-critical tasks, profiling consumes over 99\% of the planning time. These results reinforce the importance of minimizing profiling cost. They also highlight the efficiency of our lightweight CAMO, which requires only 40-100 ms per proposing step.}


\begin{figure}[t]
  \centering
    {
    \subfigure[Resource occupancy breakdown]{\includegraphics[width=0.48\linewidth]{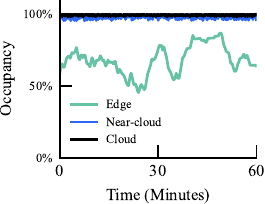}}
    \hfill
    \subfigure[Sensitivity of CAMO Committee]{\includegraphics[width=0.48\linewidth]{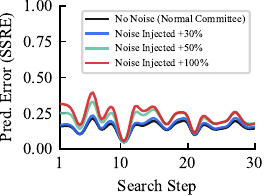}}

  }
  \caption{Resource occupancy (Left) and Sensitivity study of CAMO (Right)}
  \label{fig:occupancy-breakdown}
\end{figure}

\bipar{Breakdown by resource occupancy.} We next break down the scheduling decision. We measure \textbf{resource occupancy}, representing the percentage of queries using different tiers (\eg, 50\% of the edge occupancy means, at that time, running queries are using 50\% of edge devices). As shown in Figure~\ref{fig:occupancy-breakdown}(a), \name is able to keep near-cloud and cloud machines occupied to make sure SLO is met, while also allocating edge devices when possible, saving cloud resource to potentially accommodate more queries.




\subsection{Ablation Study}
\label{eval:ablation-study}

\begin{figure}[t]
  \centering
    {
    \subfigure[Impact of system loads. \label{fig:load}]{\includegraphics[width=0.49\linewidth]{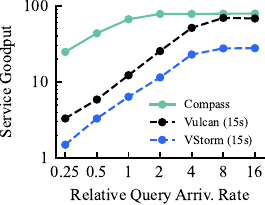}}
    \hfill
    \subfigure[Impact of planning time budget.\label{fig:cutoff}]{\includegraphics[width=0.49\linewidth]{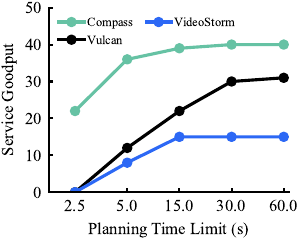}} 
    \subfigure[Impact of transfer size.
    \label{fig:abla-tf-size}]{\includegraphics[width=0.48\linewidth]{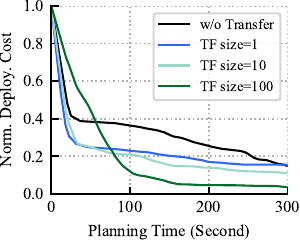}}
    \hfill
    \subfigure[\name is near-optimal. \label{fig:abla-optimal-scalable}]{\includegraphics[width=0.48\linewidth]{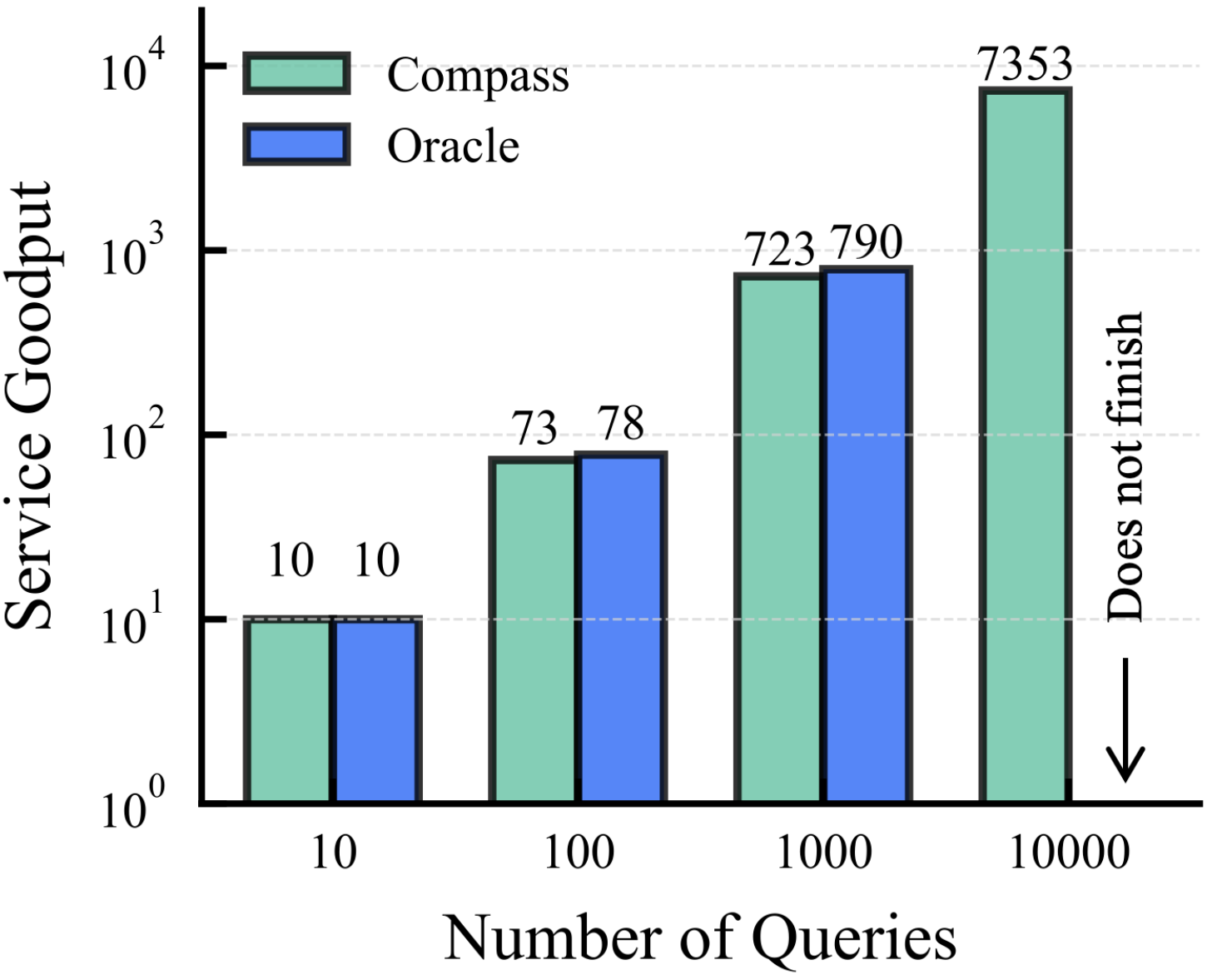}}
  }
  \caption{\name's performance in a wide range of settings.}
\end{figure}


\bipar{Sensitivity study on CAMO warm start.}

\revision{
To evaluate the impact CAMO models within the warm-start committee. We inject prediction errors into half of the historical models to simulate the presence of low-quality surrogates. For example, 30\% error means that we add an error of value 30\% of the original prediction.

As shown in Figure\ref{fig:occupancy-breakdown}(b), the committee's aggregated prediction remains stable even under such perturbations. This robustness stems from Compass's dynamic weighting mechanism: committee members are continuously reweighted based on their prediction accuracy on the new query over each step, with little overhead. Models that produce large errors are rapidly down-weighted, while better surrogates dominate the consensus, as shown in Figure~\ref{fig:abla-poison-bo-weight-x} in Appendix~\ref{app:eval_results}.
}

\bipar{Impact of system loads.}
We deploy a visual tracking task and report the service goodput to study the impact of system load (by default \textbf{1}, detailed in \S\ref{eval:setup}).
Loads $>1$ indicate overload (queuing), while $<1$ indicate underutilization. Figure~\ref{fig:load} shows that (i)~\name consistently delivers higher goodput than baselines, and (ii) as load exceeds $1$, the improvement narrows due to resource saturation (\ie, goodput capped by capacity but not \name's capability), with baselines appearing to catch up because we are sampling queries uniformly, thus it can prioritize simpler, less resource-intensive queries. (iii) the gap between \name and baselines is largest when the load is light (\ie, 0.25). We delve into that and find out that, despite having a 15s budget (3$\times$ that of Compass), baselines cannot find a feasible solution for many requests. As a result, even if the cluster is idle, baselines are not able to deliver goodput because all its plans will violate SLO. This is consistent with the result in Figure~\ref{fig:single-ttft}. 

\bipar{Impact of planning time limits.}
We evaluate \name's performance under varying planning time budgets (i.e., the responsive requirement for real-time tasks) using the live agent chat task. Figure~\ref{fig:cutoff} shows that: (i) \name consistently outperforms baselines across different planning time budgets; 
and (ii) \name achieves comparable performance to Vulcan and VideoStorm with nearly 10$\times$ less search time, making real-time serving practical in seconds. The performance gap narrows as the planning time increases because \name already identifies near-optimal plans early.

\bipar{Impact of history size.}
We evaluate the impact of the number of historical queries used in inter-query warm-start (\S\ref{sec:search-optimizer}). Figure~\ref{fig:abla-tf-size} compares settings with no transfer and with 1, 10, and 100 queries in the pool. 
Our results show that using a dozen historical models already achieves great improvements, so we use this setting as the default. Using too many (e.g., 100) introduces early-stage overhead, as \name must evaluate and select the most relevant histories, and perform hundreds of CAMO predictions.

\bipar{\name empirically achieves near-optimal performance.}
Figure~\ref{fig:abla-optimal-scalable} shows that \name closely matches the near-optimal global planning oracle (exhaustive search + ILP packing), while scaling to tens of thousands of requests. Constructing the oracle itself is extremely expensive---we spend weeks profiling all candidate plans, and the ILP solver alone takes over 4 hours for just 1K queries---impractical for online serving. We leave the scheduling scalability experiment for \name in Appendix~\ref{app:eval_results}.

\bipar{Impact of SLO requirements.}

Figure~\ref{fig:ablation-slo} shows \name's service goodput improvement under varying SLO latency tightness, and we vary the SLO accuracy tightness in Appendix~\ref{app:eval_results}. For latency SLO, we define easy, medium, and hard SLOs as 2.0$\times$, 1.5$\times$, and 1.1$\times$ the Pareto distribution mean. 
We evaluate \name on a large-scale Speech Recognition task and observe that \name achieves 2.5--5.0$\times$, 1.5--3.5$\times$ higher goodput compared to the 15s-baseline.






\bipar{Handling runtime dynamics.}
Figure \ref{fig:ablation-dynamic} shows that \name can handle runtime latency dynamics within seconds. We run an agent chat task in a scenario where the network outage drops the bandwidth between the edge and the cloud to drop to 0.2 Gbps at 26s. \name reacts swiftly, finding a new SLO-compliant plan within seconds. We left the accuracy dynamics experiments in Appendix~\ref{app:eval_results}.


\bipar{Breakdown of SLO Profiler.}
\revision{
We further break down our SLO profiler using the agentic code generation task. We disable \emph{Adaptive} profiling or \emph{Prefix Caching}, compared to the Vulcan profiler. As illustrated in Figure~\ref{fig:abla-profiler}, adaptive sampling slashes profiling time by 40\%, while prefix caching provides an additional 15\% improvement. These results demonstrate the integral effectiveness of our design.
}

\bipar{\name across deployment spaces.}
\revision{
\name's design generalizes to various deployment spaces. First, a single-tier (cloud-only) setup (Figure~\ref{fig:abla-onetier-srp}) with 4 cloud machines, corresponds to VideoStorm's design space. In the single-tier setup, VideoStorm slightly outperforms Vulcan, consistent with their design goals: VideoStorm is optimized for cloud-only deployments, while Vulcan is for edge-cloud setups.
We also show Compass's consistent improvement in a four-tier setup in the Appendix (Figure~\ref{fig:abla-fourtier-srp}).
}

%% file: sections/related.tex
\section{Related Work}
\label{sec:related}

\bipar{ML Inference Optimization.}
Optimizing ML inference efficiency has gained significant attention recently. Existing advances span model parallelism~\cite{alpaserve-osdi23, helix-asplos25}, quantization~\cite{awq-mlsys24, llmint8}, memory efficiency~\cite{vllm-sosp23}, and scheduling~\cite{vtc-osdi24, orca-osdi22} for in-cluster model serving~\cite{powerinfer-sosp24}. However, they focus on serving single ML models, or query pipelines in the cloud~\cite{zhang2017videostorm}. 
Extending these ML serving efforts toward the edge will only grow more important (\eg, distributing data preprocessing closer to the edge~\cite{recall-arxiv24}). \name complements these advancements by providing multi-tier support (\S\ref{sec:implementation}).

\bipar{Edge-cloud AI Serving.}
Recent studies have explored moving computation to the edge to improve efficiency and privacy~\cite{oort-osdi21, federated-learning}, particularly in video analytics~\cite{bhardwaj2022ekya}. For instance, Chameleon~\cite{jiang2018chameleon} exploits spatiotemporal correlations in video streams to achieve optimal resource-accuracy tradeoffs. EFL~\cite{efl-mobicom21} partitions video clips and then dynamically assigns these partitions to edge servers. Similarly, JellyBean~\cite{wu2022jellybean} and Vulcan~\cite{zhang2024vulcan} optimize the placement and execution of cost-efficient pipeline operators across infrastructure tiers but are restricted to handling single queries.
\name serves for generic, compound AI workloads where multiple queries share infrastructure resources, outperforming existing advances (\S\ref{sec:eval}).

\bipar{Efficient Configuration Search.}
Efficient configuration search has been extensively studied in the system domain, such as database ~\cite{dbtune-sigmod17}, cloud hardware ~\cite{cherrypick-nsdi17, ernest-nsdi16}, or ML hyper-parameter tuning~\cite{boparam-icml15}. 
Selecta~\cite{selecta-atc18} applies collaborative filtering to identify optimal cloud machine configurations. VideoStorm~\cite{zhang2017videostorm}  searches query configurations and resource demands in cloud environments. Many learned algorithms have been used to facilitate the search process, such as Bayesian optimization~\cite{zhang2024vulcan,cherrypick-nsdi17, snoek2012practical-bo-ecs}, greedy hill climbing~\cite{zhang2017videostorm}, and multi-armed bandit~\cite{hill2017efficient-multarmedbandit-ecs}. We propose a novel CAMO optimizer to tackle the considerably large search space for many queries.

%% file: sections/conclusion.tex
\section{Conclusion}
\label{sec:outtro}

This paper presents \name, a novel SLO-aware query planner for large-scale compound AI serving deployment. By introducing a two-stage planning framework and leveraging plan similarity with precision-aware profiling, \name efficiently determines operator placements, configurations, and resource allocations for many queries with diverse SLO requirements, system, and data characteristics. Our evaluations using real-world workloads show that \name improves service goodput by 2.4--5.1$\times$ and reduces deployment costs by 3.8--4.5$\times$, enabling real-time compound AI deployments.

%% file: appendix/appendix.tex
\appendix
\raggedbottom

{\color{purple}
\begin{table}[ht]
  \centering
  \small
  \begin{tabular}{|l|l|}
      \hline
      \textbf{Symbol} &\textbf{Definition} \\
      \hline
     $plan$ or $pl$ & plan of ($p$, $c$,$r$) tuple \\ 
     $p$ & placement, \ie edge or cloud \\
      $c$ & configuration knob, \ie 1B or 8B model \\
    $r$ & resource allocation, \ie 25\% or 100\% A100 \\
    $A_{slo}$ & accuracy SLO, \ie 0.8 \\
    $L_{slo}$ & latency SLO, \ie 200ms \\
    $f_l$ or $f_L$ & BO model for latency prediction \\
    $f_a$ or $f_A$ & BO model for accuracy prediction\\
    $U(plan)$ & Utility function \\
    $\textbf{C}(plan)$ & Profiling cost function\\
    $\mathcal{D}$ & dataset (distribution) \\
    $\mathcal{C}$ & All configurations \\
    $\mathcal{C_r}$ & All configuration, less searched \\
    
    \hline
  \end{tabular}
  \caption{Symbols and their definition used in the paper}
  \label{tab:all_symbols}
\end{table}
}

\input{appendix/setup}
\input{appendix/mp_proof}

\input{appendix/latency}
\input{appendix/discussion}
\input{appendix/additional_results}

%% file: appendix/setup.tex
\section{Evaluation Setup}
\label{appendix:setup}
\subsection{Tasks}
\begin{denseitemize}
    \item \textit{Visual Tracking}: Given a video clip, the system tracks the object of interest. We use MOT17, MOT20, GMOT-40, DanceTrack, SportsMOT and VisDrone ~\cite{mot17ds, mot20ds, gmotds, dancetrackds, sportsmotds, visdroneds} datasets. The pipeline consists of a sampler, an augmenter, a detector, and a tracker. 
    
    
    \item \textit{Speech Recognition}: Converts live human speech with background noise into text. We use FLoRes, VOiCES, MInDS-14, LibriSpeech, TED-LIUM, and VoxPopuli ~\cite{voicesds, floresds, mindsds, librids, liumds, voxds} datasets. The pipeline consists of a sampler, a denoiser, an encoder, and a decoder. 
    
    
    \item \textit{Live Agent Chat}: Given a video clip and a multi-choice question, the system generates an answer. We use NExT-QA, SUTD-TrafficQA, and Video-MME ~\cite{nextqads,sutdds,videommeds} datasets. The pipeline consists of a sampler, an image processor, and a multi-modality model. 
    
    
    \item \textit{Video Captioning}: Given a video clip, the system temporally segments it into chapters. We use VidChapters-7M, Video Timeline Tags (ViTT), YouCook2, and ActivityNet ~\cite{yang2023vidchapters,huang2020multimodal,ZhXuCoCVPR18,krishna2017dense} datasets. The pipeline consists of an ASR extractor, a video sampler, a vision backbone and a pretrained Vid2Seq~\cite{yang2023vid2seq} model. 

    \item \textit{Agentic Code Generation}: Given a natural language description of a task, the system generates a program. We use HumanEval, MBPP, and DS-1000 ~\cite{humanevalds, mbppds, ds1000ds} datasets. The pipeline consists of an analyzer and a code generator.

    
\end{denseitemize}

The datasets' realistic raw lifespan and data distribution are used, with queries lasting from a few to tens of minutes. 

\subsection{Cluster Setup}

\begin{itemize}
    \item One-tier setup: It only consists of the cloud tier, with A100 80 GB GPUs.
    \item Three-tier setup: It consists of cloud tier, near-cloud tier and edge tier.
    \item Four-tier setup: It consists of cloud tier, near-cloud tier, near-edge tier and edge tier.
\end{itemize}

For edge devices, we conduct profiling on commodity GPUs, including Nvidia RTX 1060, 1080Ti, 2080, and 3090 GPUs on VastAI~\cite{vastaiplatform}. Their CPU and memory setup varies, involving Intel Core i9, i10, and AMD Ryzen, EPYC CPUs, and DRAM in a range of 12-64GB. With our profiled result, we further diversify the edge devices using a Mobi~\cite{mobiperf-url} trace consisting of 10 million edge devices, including SoCs.

For network setup, we use provisioned bandwidth for the inter-cluster network. For the intra-cluster network, we set the edge to cluster upload bandwidth to be 50 Mbps, mimicking WIFI. For rest networks (including edge download and other intra-cluster networks), we set bandwidth to 500 Mbps, mimicking WAN speed.

\subsection{Knobs}

\begin{itemize}
  \item \textbf{Speech Recognition}
  \begin{itemize}
    \item \textbf{Input Sample Rate:} 10k, 12k, 14k, 16k
    \item \textbf{Denoiser Frequency Mask Width:} 0, 50, 100, 200
    \item \textbf{Inference Model:} wav2vec-base, wav2vec-large-10m, wav2vec-base-960h, hubert-large, hubert-xlarge
  \end{itemize}

  \item \textbf{Visual Tracking}
  \begin{itemize}
    \item \textbf{Input Frame Size:} $1280 \times 1280$, $640 \times 640$, $416 \times 416$
    \item \textbf{Detection Model:} YOLOv8x, YOLOv8n, YOLOv8s, YOLOv8m
    \item \textbf{ReID Model:}\\
    ResNet152, ResNet101, ResNet50, ResNet18, MobileNet
  \end{itemize}

  \item \textbf{Live Agent Chat}
  \begin{itemize}
    \item \textbf{Number of Frames:} 2, 4, 6, 8, 10
    \item \textbf{Image Resolution:} 256, 448, 896
    \item \textbf{Multi-Modality Model:} InternVL-1B, InternVL-2B, InternVL-4B, InternVL-8B, Gemma-3B, Gemma-27B
    \item \textbf{Speech to Text and Text to Speech:} No additional knobs.
  \end{itemize}

  \item \textbf{Video Captioning}
  \begin{itemize}
    \item \textbf{Input Sample Rate:} 0.5, 1.0, 1.5, 2.0
    \item \textbf{Image Resolution:} 224, 446
    \item \textbf{Vision Encoder:}\\
    ViT-L-14-336 (fp32, mixed),\\
    ViT-B-16-SigLIP (fp32, fp16-i18n-256),\\
    convnext\_large\_d (fp16)
    \item \textbf{Transcript Extractor:} Whisper-large-v2, Whisper-medium, Whisper-small
    \item \textbf{Vid2Seq Checkpoint:} 
    \begin{itemize}
      \item HowTo100M + VidChapters-7M
      \item HowTo100M + VidChapters-7M + YouCook2
      \item HowTo100M + VidChapters-7M + ViTT
    \end{itemize}
  \end{itemize}

\revision{
\item \textbf{Agentic Code Generation}
  \begin{itemize}
    \item \textbf{Analyzer:}\\
    LLama3.2-1B, LLama3.2-3B, Qwen2.5-0.5B, Qwen2.5-1.5B, Qwen2.5-3B, Qwen2.5-7B, Qwen-2.5-14B, Qwen-2.5-32B
    \item \textbf{Code Generator:}\\
    Qwen2.5-0.5B, Qwen2.5-1.5B, Qwen2.5-3B, Qwen2.5-7B, Deepseek-Coder-1.3B, Deepseek-V2-Lite.
  \end{itemize}
\end{itemize}
}

\subsection{Addtional Notes}

\begin{center}
    \includegraphics[width=0.6\linewidth]{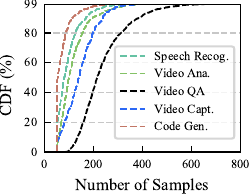}
    \captionof{figure}{C.D.F of number of samples for adaptive profiling.}
    \label{appendix:fig:ttest-cdf}
\end{center}

\bipar{Adpative Profiling} To ensure fair comparison, we set the number of samples w/o adaptive sampling to be the P99 number derived from t-test, in Figure \ref{appendix:fig:ttest-cdf}. Here we report the number used in evaluation: SR (356), VT (353), LVC (688), DVC (365), and ACG (252).

Given that the t-test requires normal distribution, we set the minimum number of samples in the guided sample to be 50, in order to have a sample size to use the central limit theorem.

\bipar{Prototype Implementation}
\revision{
We implement the prototype system atop k8s. The table below shows the overhead from our implementation. We also provide the script to measure the k8s overhead.
}

\begin{table*}[h!]
\begin{tabular}{ccccc}
\hline
\multicolumn{1}{|c|}{\multirow{2}{*}{K8s Setup}} & \multicolumn{1}{c|}{\multirow{2}{*}{gRPC Connection Setup}} & \multicolumn{3}{c|}{Load Model / States}                                                                          \\ \cline{3-5} 
\multicolumn{1}{|c|}{}                     & \multicolumn{1}{c|}{}                      & \multicolumn{1}{c|}{non-LLM}        & \multicolumn{1}{c|}{LLM ($\sim$1.7B)} & \multicolumn{1}{c|}{LLM ($\sim$7B)} \\ \hline
\multicolumn{1}{|c|}{1.1s}                 & \multicolumn{1}{c|}{< 0.1s}                  & \multicolumn{1}{c|}{< 0.1s} & \multicolumn{1}{c|}{0.2s}             & \multicolumn{1}{c|}{0.6s}           \\ \hline
\multicolumn{1}{l}{}                       & \multicolumn{1}{l}{}                       & \multicolumn{1}{l}{}                & \multicolumn{1}{l}{}                  & \multicolumn{1}{l}{}               
\end{tabular}
\caption{A detailed breakdown on migration overhead. We discuss the detail of k8s setup in question 3.}
\label{tab:migration-breakdown}
\end{table*}

In multi-query migration, these setup and loading steps execute in parallel in our prototype, so the overhead does not accumulate across migrated queries. \name is also complementary to migration optimizations such as streaming state transfer, which could further reduce adaptation latency.

\begin{lstlisting}[breaklines=true,columns=fullflexible]
kubectl delete node "$K8S_NODE" --ignore-not-found=true
sudo ssh -o StrictHostKeyChecking=no "$SSH_HOST" "sudo kubeadm reset -f 
&& sudo rm -rf /var/lib/cni/ 
&& sudo ip link delete cni0 2>/dev/null || true 
&& sudo ip link delete flannel.1 2>/dev/null || true 
&& sudo systemctl restart containerd"

START_TIME=$(date +%s.%N)

JOIN_CMD=$(sudo kubeadm token create --print-join-command)
sudo ssh -o StrictHostKeyChecking=no "$SSH_HOST" "sudo $JOIN_CMD" &
while ! kubectl get node "$K8S_NODE" >/dev/null 2>&1; do
     sleep 0.1
done

END_TIME=$(date +%s.%N)
DURATION=$(awk "BEGIN {printf \"%.6f\", $END_TIME - $START_TIME}")
\end{lstlisting}

\newpage

%% file: appendix/mp_proof.tex
\section{Theoretical Guarantee}
{\color{black}
\label{appendix:mp-proof}

\begin{highlighttheorem}
\begin{theorem}
Let $A(t)$ and $C(t)$ denote the goodput and deployment cost achieved by our greedy planner at time~$t$, and let $A_{\mathrm{opt}}(t)$ and $C_{\mathrm{opt}}(t)$ denote the corresponding optimal values. Then:
\begin{denseenum}
    \item In cloud-only setups, $A(t) \geq \frac{1}{2} A_{opt}(t)$; 
    \item Let $d$ be the number of tiers, $C(t)\leq \frac{11}{9}\times C_{opt}(t) + O(d)$.
\end{denseenum}
\end{theorem}
\end{highlighttheorem}

In the theorem, $A(t)$ and $A_{opt}(t)$ correspond to the case of limited resources, where we optimize for maximum goodput. $C(t)$ and $C_{opt}(t)$ correspond to the case of unlimited resources, where we optimizes for accomodate all queries with a minimal amount of resources.

\bipar{Notations} Let $\mathcal{Q}=\{Q_1, Q_2...Q_n\}$ be the incoming queries, for each query $Q_i\in\mathcal{Q}$. The single query search phase has generated a set of Pareto-optimal plans $C_i=\{c^{i}_, c^{i}_2 ...\}$. Each plan $c^i_j$ is a resource cost vector $\Vec{c^i_j[m]}=(t_1, t_2...t_m)$ across all $m$ hardware types (tiers). 

\subsection{Proof of (1)}
\revision{
\bipar{Outline} We prove (1) by mapping the multi-query scheduling problem to MDMCK (Multi-Dimensional Multiple-Choice Knapsack) Problem. Then we show that, in the cloud-only setup, it becomes the basic 1-dimension bin packing problem. Then we show that \name greedy is equivalent to First Fit Decreasing, and invoke the bound from previous literature to get the $\frac{1}{2}$ factor. 

\bipar{Problem Formulation}

When the resource is limited, our goal is to accommodate as many queries as possible. We assume there are $L_m$ GPUs available on tier $m$, each with capacity $T_m$. We also assume the weight of query $Q_i$ is $W_i$. The optimization goal is to maximize the goodput of queries we can accommodate.

\bipar{Decision Variables} We introduce a binary variable $h_i$ indicating whether we decide to admit the query $Q_i$.We have $x_{ij}$ is a binary variable indicating whether we use the plan choice $c_j^i$ for query $Q_i$. $y_{ijk}$ is a binary variable indicating whether the placement choice $c_j$ for $Q_i$ is placed on $k$th GPU. 

\begin{align}
    maximize: & \sum_{s=1}^{N} W_s h_s \\
s.t. & \sum_{i,j} h_i x_{ij} y_{ijk}^m c_j^i[m] \leq T_m, \forall m\in [1, M], k \in [1, L_m] \\
&\sum_{j=1}^{|C_i|} x_{ij} = h_i, \forall i \in [1, N]  \\
&\sum_{k=1}^{N} y_{ijk}^m = h_i, \forall i \in [1, N], j \in [1, |C_i|], m \in [1, M]\\
&h_{i} \in \{0,1\}, x_{ij} \in \{0,1\}, y_{ijk} \in \{0,1\} 
\end{align}

There are $O(N*M*|C|*L)$ binary variables and $O(N*M*|C|+ M * L)$ constrains.

\bipar{Mapping}
The problem is harder than MDMCK because we also choose which GPU to use. It is not only NP-hard but also APX-hard~\cite{chekuri2004multidimensional-apxhard}. Thus we cannot guarantee any polynomial-time algorithm when $m>1$.

\bipar{Cloud-only Case} In the cloud-only setup, we have $m=1$. $m=1$ also implies there is only one plan in the Pareto-frontier (\ie, $C_i=\{c^i\}$). Thus, we can simplify the formulation into:
\begin{align}
    maximize: & \sum_{s=1}^{N} W_s h_s \\
s.t. & \sum_{i,j} h_i y_{ik} c^i \leq T_m, \forall k \in [1, L] \\
&\sum_{k=1}^{N} y_{ik} = h_i, \forall i \in [1, N]\\
&h_{i} \in \{0,1\}, y_{ik} \in \{0,1\} 
\end{align}

Note that we are only making two decisions: $h_i$ is whether we accommodate one $i$, and $y_{ik}$ is which GPU to use for query $i$. We have items of different sizes (query plan with different resource cost and goodput gain), and containers with the same capacity (GPUs). This is 1-dimension bin packing problem. 

\bipar{Greedy Solution}
\name greedy algorithm sorts queries by unit goodput gain per monetary cost, using a weighted average of resources on each tier. It then accommodates as many queries as possible in that order. When $m=1$, the sorting criterion becomes unit goodput per resource cost, equivalent to First Fit Decreasing (FFD) with smallest-first ordering. FFD provides a 2-approximation~\cite{coffman1984approximation-binpacking-2-approximate}
, i.e., the worst-case scenario is at most half of the goodput from the optimal, thus completing the proof where $A(t) \geq \frac{1}{2}A_{opt}(t)$.
}

\subsection{Proof of (2)}
\revision{
\bipar{Outline} We prove (2) by first mapping the problem into $m$ independent 1-dimensional bin packing problems. Then we show that \name greedy is equivalent to the First Fit Decreasing strategy. Then we invoke the bound from previous literature to get the $\frac{11}{9}$ factor with a constant. At last, we prove that the extra cost incurred from fragmentation can be accounted for by a small constant.

\bipar{Formulation} 
In this case, the optimization objective is to accommodate all queries with minimum monetary cost. We define the monetary cost of tier (hardware) $m$ as $G_m$.  Assume we use $K_m$ GPU on tier $m$, our goal is to minimize the monetary cost $\$ = \sum K_m * G_m$.

\bipar{Decision Variables}

We have $x_{ij}$ is a binary variable indicating whether we use the plan choice $c_j^i$ for query $Q_i$. $y_{ijk}^m$ is a binary variable indicating whether the placement choice $c_j$ for $Q_i$ is placed on $k$th GPU at tier $m$. $z_k^m$ is a binary variable indicating whether the $k$th GPU at tier $m$ is used; assume we use at most $K$ GPUs (the cost where all queries uses most costly plan). $T$ is the resource capacity for a single GPU. 

\begin{align}
    minimize: & \sum_{t=1}^{m}\sum_{s=1}^{K} G_t * z_s^m \\
s.t. & \sum_{i,j} x_{ij} * y_{ijk}^m * c_j^i[m] \leq T_m * z_k^m, \forall m \in [1, M], k\in [1, K] \\
&\sum_{j=1}^{|C_i|} x_{ij} = 1, \forall i \in [1, N]  \\
&\sum_{k=1}^{N} y_{ijk}^m = 1, \forall i \in [1, N], j \in [1, |C_i|], m \in [1, M]\\
&x_{ij} \in \{0,1\}, y_{ijk} \in \{0,1\}, z_k \in \{0,1\} 
\end{align}

There are $O(N*M*|C|*K)$ binary variables and $O(N*M*|C|+ M * K)$ constrains.

\bipar{Mapping} Our goal is to map this problem into $m$ independent 1-dimensional bin-packing problems. However, the current formulation is not independent, because we need to select plan $c_i^j$. To solve this problem, we first assume no fragmentation. In that case, given that we have unlimited resources and the optimization goal, we are always going to choose the plan with the smallest monetary cost. Lets denode $c_0^i$ be that one.

\begin{align}
    minimize: & \sum_{t=1}^{m}\sum_{s=1}^{K} G_t * z_s^m \\
s.t. & \sum_{i,j} y_{ik}^m * c_0^i[m] \leq T_m * z_k^m, \forall m \in [1, M] k\in [1, K] \\
&\sum_{k=1}^{N} y_{ik}^m = 1, \forall i \in [1, N], m \in [1, M]\\
& y_{ik} \in \{0,1\}, z_k \in \{0,1\} 
\end{align}

After the reduction, the decision variables of $m$ tiers are independent of each other. Instead of solving an m-dimensional bin-packing problem, it becomes m 1-dimensional bin-packing problems.

\bipar{Greedy Solution}

\name greedy algorithm sorts the query based on unit goodput gain per monetary cost (a weighted average of resource on each tier). Then \name sort queries based on their unit gain, and try to accommodate as many as possible based on the order. This is equivalent to that of First Fit Decreasing (FFD) with smallest-first ordering. When considering $m$ together, let the cost of herustic be $ALG$ and optimal value be $OPT$, it is proved that $ALG \leq \alpha OPT + d\beta$, where $\alpha=\frac{11}{9}$ and $\beta$ is a small constant, i.e., $O(m)$~\cite{johnson1974worst-11-9}.

\bipar{Fragmentation}

We then show that the potential cost from fragmentation is less than $O(m)$, by showing that for each tier, it would be less than $O(1)$. This comes straightforwardly from our quantization method; We only allow values to be a multiple of the divisor of 100\% (For example, in A100 we allow 25\%, 50\%, 75\%, 100\%). This is Bin Packing with Divisible Item Sizes, and it is proven that the cost is $\left\lceil\frac{\text{total size}}{\text{bin size}}\right\rceil$~\cite{coffman1987binpackingdivisible}. Thus, we are wasting at most $O(1)$ GPU. Given that $O(m)+O(m)=O(m)$, we make sure the constant is accounted for.

Putting everything together, if we use $d$ to denote number of tiers, we have $C(t)\leq \frac{11}{9}C_{opt}(t)+O(d)$.
}
}

%% file: appendix/latency.tex
\section{\name Interface}
\label{app:interface}

Figure~\ref{code:interface} illustrates an example of \name interface, which complements existing ML pipelines with only a few lines of code changes in their APIs. 

\begin{figure}[t]
\begin{lstlisting}[language=rust,label={lst:interface},escapechar=|]
{  // Query example
  "stages": [["Analyzer", "Generator"], 
  ["Detector", "Cloud"],["Generator", "Cloud"]],
  "config": {"analyzer": "Llama3.2-1B","generator": "Qwen3-4B"}
}
// API example
pub async fn submit(query: Json<Request>, 
  device_pool: &mut DevicePool, controller: &mut Controller) -> Result<HttpResponse, Error> {
    // Get workers from device pool based on the query
    let workers = device_pool.take_from_stages(query.stages)?;
    // Establish gRPC connections among workers
    let channels = gRPC::join_all(workers.iter()
    .map(|w| w.connect())).await.map_err(error)?;
    // Submit worker task to the controller
    controller.add_task(channels, query.config)?;
    HttpResponse::Ok().text("Task submitted!") }

\end{lstlisting}
  \caption{An example of \name API for code generation task.}
  \label{code:interface}
\end{figure}

\section{Latency Estimation and Isolation Profiling}
\label{appendix:simulator}

The latency of a pipeline typically depends on the input length and output length (e.g., LLM decoding) or remains constant (e.g., streaming). We assume the latency in the Service Level Objective (SLO) is in a normalized form appropriate for its specific setup; however, \name is general and can be applied to any latency model.

During accuracy profiling, \name records each query's input/output length, inter-operator data movement, and per-operator computation time on a cloud A100 GPU. We model pipeline latency as the longest path in its computation graph (usually a DAG). Operator latency is computation time plus data movement time $T_{data} = \frac{L}{B} + T_0$, where $L$ is data length, $B$ is bandwidth, and $T_0$ accounts for network latency.

\revision{
\subsection{Hardware and Virtualization Mechanisms}
To estimate computation time across different GPUs and isolation environments, \name adopts the best available mechanism per platform, following existing advances~\cite{zhu2025nanoflow}. We clarify our profiling approach across different hardware:
\begin{denseitemize}
    \item \textbf{Virtualization Mechanisms:} We use Multi-Instance GPU (MIG) on A100 GPUs and Multi-Process Service (MPS) on earlier-generation devices such as V100 and T4 where MIG is unavailable. Both mechanisms are industry standards for resource isolation.
    \item \textbf{Interference Assessment:} To ensure virtualization provides acceptable isolation for planning-level modeling, we conducted experiments on RTX 3090 edge devices by co-locating two visual tracking pipelines. The observed latency overhead was minimal ($\approx 7\%$), validating the robustness of our profiling.
\end{denseitemize}

\subsection{Resource Modeling and Partitioning}
To estimate the computation time for an operator over different GPUs, we leverage state-of-the-art simulators~\cite{wang-nsdi2025simai}, analytical models~\cite{zhu2025nanoflow}, and our profiling results. We employ different resource reduction strategies based on the task type:

\paragraph{Non-batching Tasks:} 
A 25\% resource allocation means a task is allocated 25\% of the FLOPS. To further mitigate potential contention effects, \name employs \textbf{conservative resource partitioning} during profiling. Rather than relying on fine-grained fractional allocations, we restrict GPU slices to coarse levels (e.g., 25\%, 50\%, 75\%, 100\%) under MIG or MPS. This design improves robustness across hardware platforms while preserving predictability in latency estimation.

\paragraph{Batching Tasks (LLMs):} 
For LLM workloads, co-location is less common due to memory constraints. In this setting, we model "resource allocation" through \textbf{batch capacity} instead of GPU slicing. For example, a 25\% resource allocation implies the task is batched with requests from other users in a batch size of 4, while the LLM still utilizes 100\% of the FLOPS and memory. Following existing advances~\cite{zhu2025nanoflow, wang-nsdi2025simai}, we estimate latency as a function of batch size using analytical interpolation. Across our evaluations, this approach achieves high accuracy, with prediction errors within approximately 5\%, consistent with reports on ML runtime simulation~\cite{agrawal2024vidur}.

Finally, we find that the latency for inter-cluster data transfer is almost negligible compared to intra-cluster transfer given high network bandwidth. Communication overhead (e.g., from tensor parallelism) during LLM inference is treated as compute operator latency and included in the profiling stage.
}

%% file: appendix/discussion.tex
\section{Discussion}
\label{appendix:discussion}
\begin{center}[t]
    \includegraphics[width=0.6\linewidth]{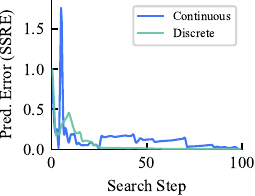}
    \captionof{figure}{Encoding configuration into discrete one-hot features sometimes outperforms encoding configuration into continuous variables for BO models in our task.}
    \label{fig:Appendix_BO}
\end{center}

\bipar{Discrete and Continuous Variables in BO}
Our comparative experiments indicate that using discrete one-hot vectors to encode configuration features can accelerate the convergence of BO and reduce inference latency, thereby enhancing overall performance.

\bipar{Data Distribution Shift.}
\revision{
Data distribution shifts are inevitable. To address this issue, we position \name as a high-speed planning component within a decoupled serving architecture. Compass integrates with an upper-level monitoring layer following previous literature~\cite{zhang2024vulcan} that tracks real-time performance signals. When significant deviations in latency or accuracy are detected, the monitor triggers an online re-planning cycle. By re-profiling on a buffered window of recent ``live'' data, \name can rapidly generate updated execution plans that reflect the current distribution. 
}

\bipar{Resource-Pareto Frontier Construction.}
The Search Optimizer assumes over-provisioned execution, such as dedicating a cloud GPU to the query, to avoid prematurely reasoning about fractional resource sharing. Fine-grained allocation is then handled by the SLO Profiler through resource trimming. Once a plan satisfies the accuracy SLO under over-provisioned resources, \name performs multi-dimensional binary search over resource dimensions. For example, it can fix the A100 allocation and reduce the V100 allocation until the resulting plan just satisfies the latency SLO. This process identifies Pareto-optimal latency-resource trade-offs without requiring combinatorial exploration during search.

\begin{figure}
    \centering
    \includegraphics[width=0.6\linewidth]{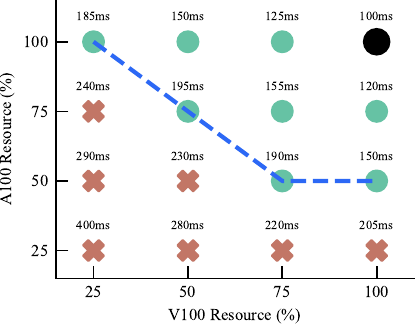}
    \caption{An example Pareto frontier with latency SLO equal to 200ms. The blue dotted line shows the Pareto frontier.}
    \label{fig:pareto-example}
\end{figure}

%% file: appendix/additional_results.tex
\section{Additional Evaluation Results}
\label{app:eval_results}

\bipar{Planning Overhead} As shown in Table~\ref{tab:total-serving}, we measure the planning time and the average serving time for three tasks. It shows that planning time is indeed the bottleneck.

\begin{table}[H]
  \centering
  \small
  \resizebox{0.99\linewidth}{!}{
 \begin{tabular}{l|l|ccc}
\hline
\multirow{2}{*}{\textbf{Task}} & \multirow{2}{*}{\textbf{Serving Time}} & \multicolumn{3}{c}{\textbf{Planning Time}} \\ \cline{3-5}
&  & \textbf{Vulcan} & \textbf{VideoStorm} & \textbf{Compass} \\ \hline
Speech Recognition & 13.8s & 20.1s & 18.9s & 3.8s  \\
Live Agent Chat & 67.2s & 30.9s & 30.8s & 5.0s \\
Visual Tracking & 37.6s & 24.2s & 26.3s & 4.4s  \\ \hline
\end{tabular}
  }
  \caption{Planning contributes a huge portion of total serving time for latency-critical requests. Serving time is measured as the average for a single request. \name significantly reduces planning time to identify the first SLO-compliant plan, enabling real-time response.}
  \label{tab:total-serving}
\end{table}

\begin{figure}[H]
  \centering
  \includegraphics[width=.45\linewidth]{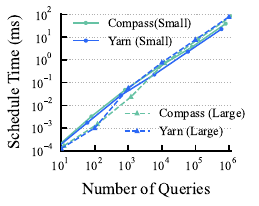}
  \caption{\name achieves strong performance and scalability.}
  \label{fig:multi-scalability}
\end{figure}

\bipar{Scheduling Scalability.}
Figure~\ref{fig:multi-scalability} compares the scheduling time between \name multi-scheduling and the Yarn scheduling with linear time complexity, which admits queries in a first-come-first-served fashion. The result shows that \name achieves similar time cost, and linearly scales up with the number of queries, in small (4 V100 + 4 A100 GPUs) and large cluster setups, respectively. ILP scheduling (not plotted) takes 4 hours to finish for 1k queries, and it does not finish after more than 24 hours for 10k queries.

\begin{figure}[H]
  \centering
  \subfigure[Online latency dynamic.\label{fig:ablation-dynamic}]{\includegraphics[width=0.45\linewidth]{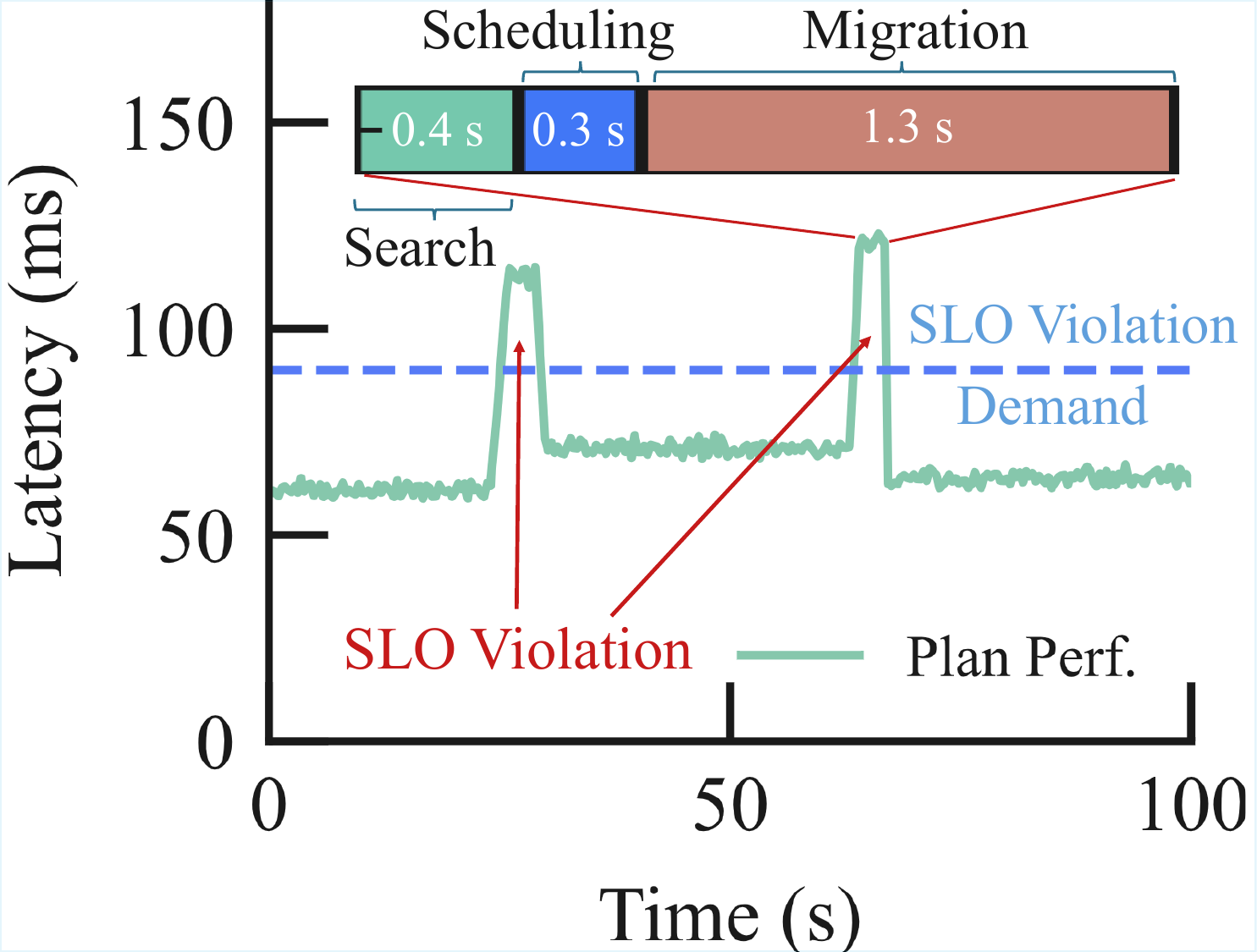}}
  \subfigure[Network Dynamic]{\includegraphics[width=0.45\linewidth]{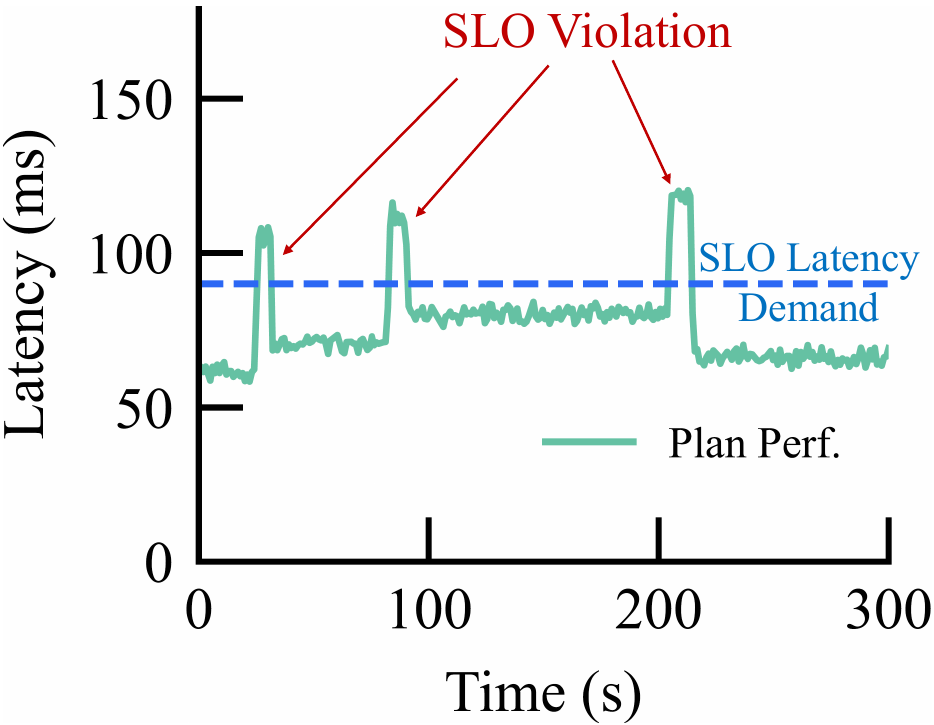}}
  \subfigure[Distribution Drift Dynamic]{\includegraphics[width=0.45\linewidth]{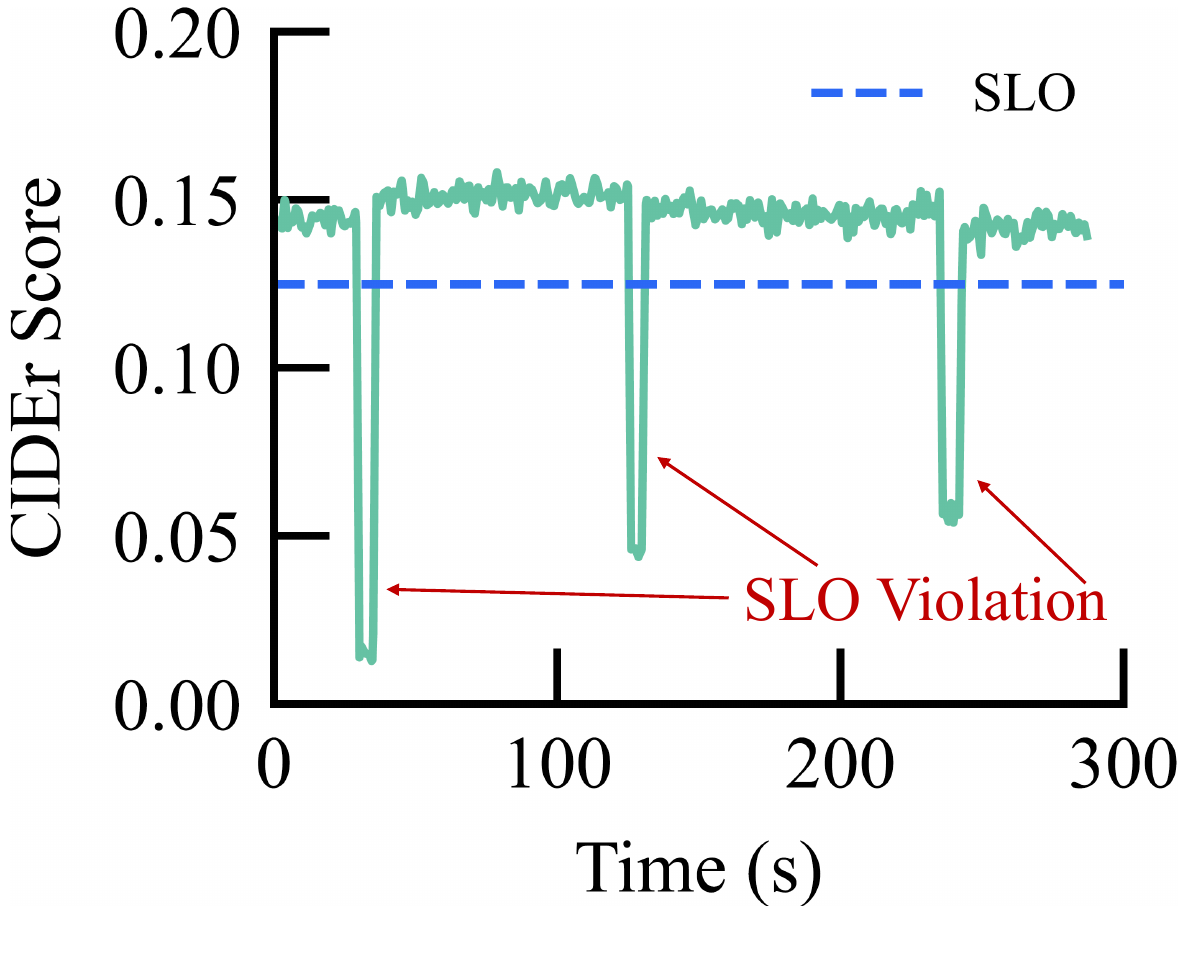}}
  \caption{\name is robust to runtime dynamics.}
  \label{fig:ablation-dynamic-app}
\end{figure}

\bipar{Handling runtime dynamics.}
Figure~\ref{fig:ablation-dynamic-app} shows that \name can handle runtime latency dynamics within seconds. We run an agent chat task in a scenario where the lighting condition of the video input changes, causing the accuracy (CIDEr score) to drop to 0.017 at 30s. \name reacts efficiently again to preserve SLOs.

\begin{figure}[H]
  \centering
  \includegraphics[width=0.45\linewidth]{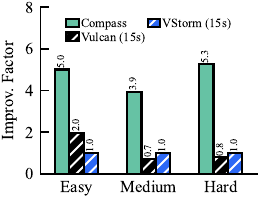}
  \caption{\name improves performance across latency SLO tightness.}
  \label{fig:ablation-slo}
\end{figure}

\begin{figure}[H]
  \centering
  \includegraphics[width=0.45\linewidth]{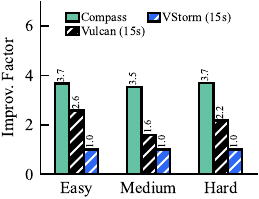}
  \caption{\name improves performance across accuracy SLO tightness.}
  \label{fig:ablation-slo-acc}
\end{figure}

\bipar{Impact of SLO requirements.}
Figure~\ref{fig:ablation-slo} shows \name's service goodput improvement under varying latency SLO tightness, while Figure~\ref{fig:ablation-slo-acc} varies accuracy SLO tightness. For the accuracy SLO, we define those as 0.7, 0.8, and 0.9 of the Pareto mean. The medium setup is used in all other experiments.
We evaluate \name on a large-scale Speech Recognition task and observe that \name achieves 2.5--5.0$\times$, 1.5--3.5$\times$ higher goodput compared to the 15s-baseline versions.

\begin{figure}[H]
  \centering
  \includegraphics[width=0.48\linewidth]{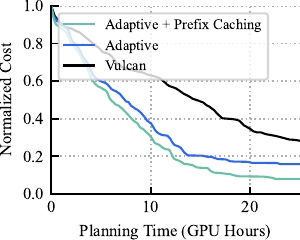}
  \caption{\name benefits from SLO profiler components.}
  \label{fig:abla-profiler}
\end{figure}

\bipar{SLO Profiler Breakdown.}
We further break down \name's SLO profiler. Figure~\ref{fig:abla-profiler} shows that \name benefits from adaptive profiling and prefix caching.

\begin{figure}[H]
  \centering
  \subfigure[Single tier cloud-native deployment.\label{fig:abla-onetier-srp}]{
    \includegraphics[width=0.48\linewidth]{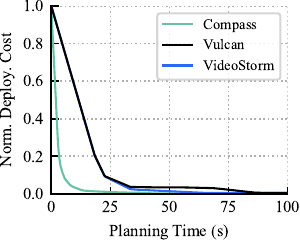}}
  \subfigure[Four-tier deployment.\label{fig:abla-fourtier-srp}]{
    \includegraphics[width=0.48\linewidth]{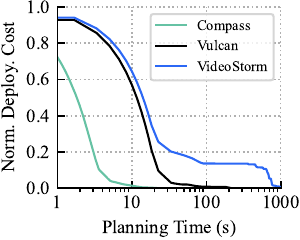}}
  \caption{\name improves across deployment spaces.}
\end{figure}

\bipar{Impact of Number of Tiers.}
We include a four-tier setup (Figure~\ref{fig:abla-fourtier-srp}), which involved all four tiers in Table~\ref {tab:cluster} by adding 16 near-edge machines with T4 GPUs to our medium-scale setup. As shown in Figure~\ref{fig:abla-fourtier-srp}, Compass achieves consistent improvement across all design spaces.

\begin{figure}[H]
  \centering
  \includegraphics[width=0.45\linewidth]{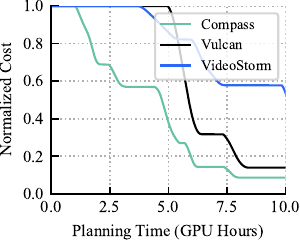}
  \caption{Agentic code generation with 14B and 32B model.}
  \label{fig:acg-ds-1000-x}
\end{figure}

\revision{
\bipar{Add Larger Model in Evaluation.}
We add 14B and 32B Qwen 2.5 models in code generation knobs, and re-run the evaluation on DS-1000. As shown in Figure~\ref{fig:acg-ds-1000-x}, Compass's advantages over baseline are consistent and even more pronounced due to its efficient profiling optimization.
}

\begin{figure}[H]
  \centering
  \includegraphics[width=0.48\linewidth]{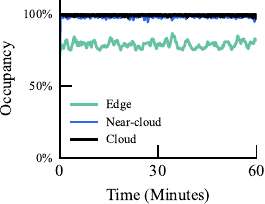}
  \caption{Occupancy breakdown on Visual Tracking Task.}
  \label{fig:occupancy-breakdown-app}
\end{figure}

\bipar{Breakdown by resource occupancy.} As shown in Figure~\ref{fig:occupancy-breakdown-app}, \name is able to keep near-cloud and cloud machines occupied to make sure SLO is met, while also allocating edge devices when possible, saving cloud resource to potentially accommodate more queries.

\begin{figure}[H]
  \centering
  \includegraphics[width=0.48\linewidth]{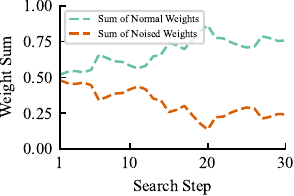}
  \caption{Sensitivity study on CAMO Committee. The figure shows the weight overtime at 50\% error.}
  \label{fig:abla-poison-bo-weight-x}
\end{figure}

\revision{
\bipar{Sensitivity Study on CAMO Committee}. As shown in Figure~\ref{fig:abla-poison-bo-weight-x}, models that produce large errors are rapidly down-weighted, while better surrogates dominate the consensus.
}

\begin{figure}[H]
  \centering
  \subfigure[Speech Recognition]{
    \includegraphics[width=0.48\linewidth]{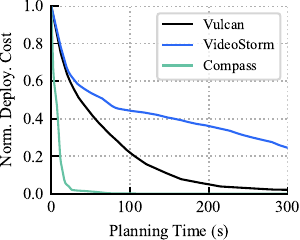}
  }
  \subfigure[Visual Tracking]{
    \includegraphics[width=0.48\linewidth]{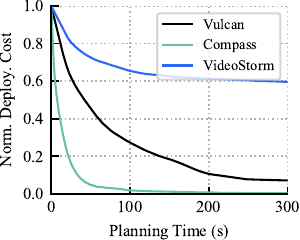}
  }
  \subfigure[Live Agent Chat]{
    \includegraphics[width=0.48\linewidth]{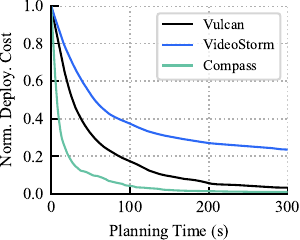}
  }
  \caption{Time-to-cost for single query.}
  \label{appendix:fig:time2cost}
\end{figure}

\bipar{Single-query Results.}
Figure~\ref{appendix:fig:time2cost} shows additional results on single-query planning, where \name uniformly outperforms baselines.

\begin{figure}[H]
  \centering
  \includegraphics[width=0.95\linewidth]{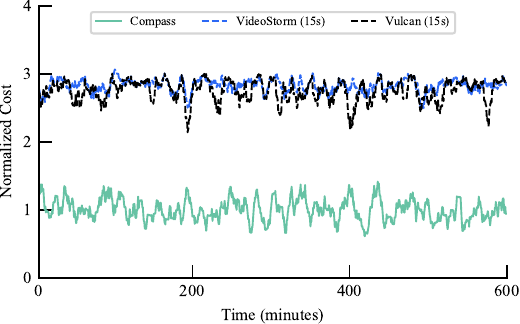}
  \caption{Live agent chat (unlimited resources).}
  \label{appendix:fig:unlimited}
\end{figure}

\begin{figure}[H]
  \centering
  \subfigure[Speech recognition]{\includegraphics[width=0.95\linewidth]{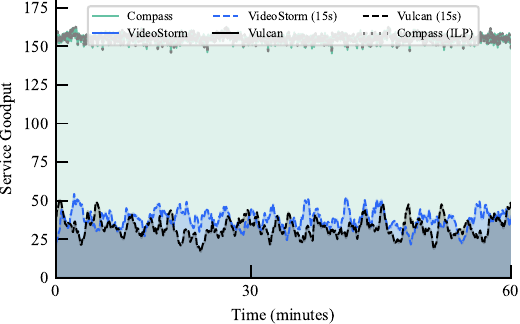}}
  \par\smallskip
  \subfigure[Visual tracking]{\includegraphics[width=0.95\linewidth]{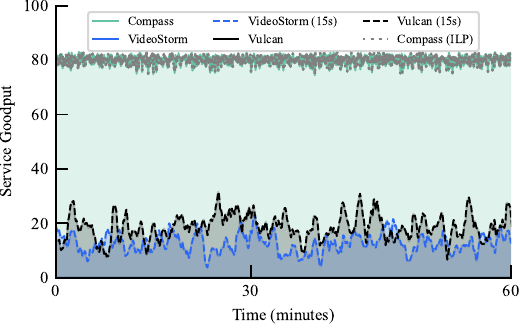}}
  \caption{Medium-scale cluster under limited resources.}
  \label{appendix:fig:small-limited-app}
\end{figure}

\begin{figure}[H]
  \centering
  \includegraphics[width=0.95\linewidth]{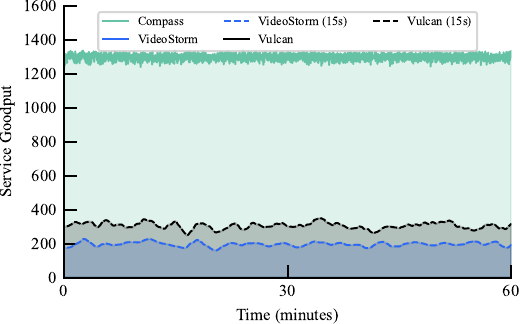}
  \caption{Visual tracking (large-scale, limited resources).}
  \label{appendix:fig:large-limited}
\end{figure}

\bipar{End-to-end Results.}
Figures~\ref{appendix:fig:unlimited}, \ref{appendix:fig:small-limited-app}, and~\ref{appendix:fig:large-limited} show additional results on end-to-end performance under different setups.

\begin{figure}[H]
  \centering
  \subfigure[Feasible rate.]{\includegraphics[width=0.49\linewidth]{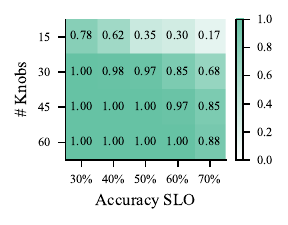}}
  \subfigure[Cost ratio Compass / Vulcan, with 5 seconds budget]{\includegraphics[width=0.49\linewidth]{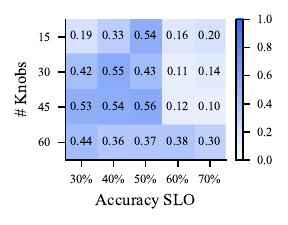}}
  \caption{We study the effect of the number of knob over different accuracy SLO. The left figure shows the rate of feasible queries, the right figure compares the cost of searched plan between \name and Vulcan.}
  \label{fig:acc_knob_heatmap-app}
\end{figure}

\revision{
\bipar{Impact of Number of Search Knobs.}
We add a new experiment using a speech recognition pipeline to systematically investigate the impact of knob dimensionality and SLO difficulty. In this setup, we vary the number of configuration knobs across (15, 30, 45, 60) and set the accuracy SLO requirements to (30\%, 40\%, 50\%, 60\%, 70\%).

\begin{denseitemize}

\item \emph{Feasibility Rate}: the fraction of queries that admit at least one SLO-compliant plan when responsiveness constraints are removed. This metric captures the intrinsic capacity of a given knob space and reflects how expressive the configuration space is for satisfying diverse SLO requirements.

\item \emph{Cost Efficiency}: the deployment cost of the identified plan under a strict search time budget (5 seconds in this experiment). This metric evaluates the practical effectiveness of different planning methods under realistic responsiveness constraints.
\end{denseitemize}

As shown in Figure~\ref{fig:acc_knob_heatmap-app}, the left heatmap demonstrates that the feasibility rate increases with the number of knobs. Intuitively, a larger knob space provides greater flexibility to satisfy stringent accuracy SLOs, enabling the system to serve more difficult queries. This result highlights both the necessity of exposing richer configuration knobs and the challenge of efficiently navigating the resulting large search space.

The right heatmap evaluates cost efficiency under a fixed 5-second search budget. \name consistently outperforms the baseline Vulcan across all settings. Notably, \name can identify plans with as low as 11\% of the deployment cost of Vulcan within the same responsiveness constraint. These results collectively validate the importance of scalable, SLO-aware planning in high-dimensional configuration spaces.
}

